\documentclass[aps,prd,amsmath,floats,floatfix,superscriptaddress,nofootinbib,
showpacs]{revtex4}

\usepackage{amssymb}
\usepackage{amsmath}
\usepackage{verbatim}
\usepackage{mathrsfs}
\usepackage{amsfonts}
\usepackage{latexsym}
\usepackage{epsfig}
\usepackage{color}
\usepackage{graphicx,subfigure}
\usepackage{units}
\usepackage{overpic}
\usepackage[hidelinks]{hyperref}
\usepackage{afterpage}
\usepackage{float}
\usepackage{mathtools}
\usepackage{xcolor}
\hypersetup{
    colorlinks,
    linkcolor={red!50!black},
    citecolor={blue!50!black},
    urlcolor={blue!80!black}
}

\usepackage{booktabs, multirow} 
\usepackage{soul}
\usepackage{changepage,threeparttable} 
\begin{document}


\definecolor{orange}{rgb}{0.9,0.45,0} 
\definecolor{applegreen}{rgb}{0.55, 0.71, 0.0}

\newcommand{\argelia}[1]{\textcolor{red}{{\bf Argelia: #1}}}
\newcommand{\dario}[1]{\textcolor{red}{{\bf Dario: #1}}}
\newcommand{\juanc}[1]{\textcolor{olive}{{\bf JC: #1}}}
\newcommand{\juan}[1]{\textcolor{cyan}{{\bf Juan B: #1}}}
\newcommand{\alberto}[1]{\textcolor{blue}{{\bf Alberto: #1}}}
\newcommand{\miguela}[1]{\textcolor{red}{{\bf MiguelA: #1}}}
\newcommand{\mm}[1]{\textcolor{orange}{{\bf MM: #1}}}
\newcommand{\OS}[1]{\textcolor{blue}{{\bf OS: #1}}}
\newcommand{\victor}[1]{\textcolor{applegreen}{{\bf Victor: #1}}}


\title{Extreme \texorpdfstring{$\ell$}{l}-boson stars}

\author{Miguel Alcubierre}
\affiliation{Instituto de Ciencias Nucleares, Universidad Nacional Aut\'onoma de M\'exico,
Circuito Exterior C.U., A.P. 70-543, M\'exico D.F. 04510, M\'exico}

\author{Juan Barranco}
\affiliation{Departamento de F\'isica, Divisi\'on de Ciencias e Ingenier\'ias,
Campus Le\'on, Universidad de Guanajuato, Le\'on 37150, M\'exico}

\author{Argelia Bernal}
\affiliation{Departamento de F\'isica, Divisi\'on de Ciencias e Ingenier\'ias,
Campus Le\'on, Universidad de Guanajuato, Le\'on 37150, M\'exico}

\author{Juan Carlos Degollado}
\affiliation{Instituto de Ciencias F\'isicas, Universidad Nacional Aut\'onoma de M\'exico,
Apdo. Postal 48-3, 62251, Cuernavaca, Morelos, M\'exico}

\author{Alberto Diez-Tejedor}
\affiliation{Departamento de F\'isica, Divisi\'on de Ciencias e Ingenier\'ias,
Campus Le\'on, Universidad de Guanajuato, Le\'on 37150, M\'exico}

\author{V\'ictor Jaramillo}
\affiliation{Instituto de Ciencias Nucleares, Universidad Nacional Aut\'onoma de M\'exico,
Circuito Exterior C.U., A.P. 70-543, M\'exico D.F. 04510, M\'exico}

\author{Miguel Megevand}
\affiliation{Instituto de F\'isica Enrique Gaviola, CONICET. Ciudad Universitaria, 5000 C\'ordoba, Argentina}

\author{Dar\'io N\'u\~nez}
\affiliation{Instituto de Ciencias Nucleares, Universidad Nacional Aut\'onoma de M\'exico,
Circuito Exterior C.U., A.P. 70-543, M\'exico D.F. 04510, M\'exico}

\author{Olivier Sarbach}
\affiliation{Instituto de F\'isica y Matem\'aticas, Universidad Michoacana de San Nicol\'as de Hidalgo,
Edificio C-3, Ciudad Universitaria, 58040 Morelia, Michoac\'an, M\'exico}


\date{\today}


\begin{abstract}
A new class of complex scalar field objects, which generalize the well 
known boson stars, was recently found as solutions to the Einstein-Klein-Gordon 
system. The generalization consists in incorporating some of the effects of 
angular momentum, while still maintaining the spacetime's spherical symmetry. 
These new solutions depend on an (integer) angular parameter $\ell$, and hence 
were named $\ell$-boson stars. Like the standard $\ell=0$ boson stars these 
configurations admit a stable branch in the solution space; however, contrary to 
them they have a morphology that presents a shell-like structure with a ``hole'' 
in the internal region. In this article we  perform a thorough exploration of 
the parameter space, concentrating particularly on the extreme cases with large 
values of $\ell$. We show that the shells grow in size with the angular 
parameter, doing so linearly for large values, with the size growing faster than 
the thickness. Their mass also increases with $\ell$, but in such a way that 
their compactness, while also growing monotonically, converges to a finite value 
corresponding to about one half of the Buchdahl limit for stable configurations. Furthermore, we show that 
$\ell$-boson stars can be highly anisotropic, with the radial pressure 
diminishing relative to the tangential pressure for large $\ell$, reducing 
asymptotically to zero, and with the maximum density also approaching zero. We 
show that these properties can be understood by analyzing the asymptotic limit 
$\ell\rightarrow\infty$ of the field equations and their solutions. We also 
analyze the existence and characteristics of both timelike and null circular 
orbits, especially for very compact solutions.
\end{abstract}




\maketitle

\section{Introduction}
\label{Sec:Intro}

The possibility that dark matter can be described by a scalar field has
recently found an increasing interest, either through the study of models with a particle physics motivation~\cite{Preskill:1982cy,Abbott:1982af,Dine:1982ah,ParticleDataGroup:2020ssz}, or through the description of lighter fields with the potential to alleviate some possible tensions in the standard cosmological scenario on small 
scales~\cite{Marsh:2015xka,Hui:2016ltb,Niemeyer:2019aqm,Hu:2000ke,Matos:1999et,Matos:2000ss,Matos:2003pe} (see e.g.~\cite{genina2018core,Kendall:2019fep,Kim:2017iwr,DES:2020fxi} 
for updated discussions on the classical cold dark matter problems). Gravitationally bound bosonic structures appearing as the consequence
of these fields may be relevant in astrophysics, as they could develop dark matter 
halos and/or very compact objects, depending on the particular choice of the parameters of the model.
In the high compactness regime, bosonic structures can approach the Buchdahl
limit~\cite{Buchdahl:1959zz}
and form objects similar in size and mass to neutron stars or even black holes.
Like other compact objects~\cite{Cardoso:2019rvt}, boson stars~\cite{Jetzer92,Liddle:1992fmk,Mielke:1997re,Liebling_2017,visinelli2021boson} may form bound binary systems 
emitting gravitational waves of distinctive features.
The dynamics of these systems has been studied for instance in
references~\cite{Palenzuela:2017kcg,Bezares:2017mzk,Bezares:2018qwa} (see also~\cite{Bustillo:2020syj} where the waveforms calculated from the head-on collision between two Proca stars is confronted with gravitational
wave observations).
On the other hand, in the low compactness regime, gravitationally 
bound structures can be used to describe dark matter halos~\cite{Sin:1992bg,Lee:1995af,Arbey:2003sj}, although some controversies have arisen regarding the
non-compatibility on the required values of the field mass when combining different data sets. 
For example, the characteristic masses needed to describe the internal kinematics of the Milky Way dwarf spheroidal satellites are in tension when faced with 
cosmology~\cite{Gonzales-Morales:2016mkl} (see also~\cite{Hayashi:2021xxu}), 
and even at local scales the mass density profiles of dwarf spheroidal and ultra-faint dwarf galaxies suggest different values for the field mass~\cite{Hayashi:2021xxu,Safarzadeh:2019sre}.
Furthermore, for larger galaxies the dark matter halos
could be even more cuspy than the standard cold dark matter
Navarro-Frenk-White profiles~\cite{Robles:2018fur}. These problems emerge when fitting the observations to the dark matter halo profile predicted by a standard boson star, in some cases enlarged with an external Navarro-Frenk-White profile 
as suggested by numerical cosmological simulations~\cite{Schive:2014dra,Schive:2014hza,Schwabe:2016rze,Veltmaat:2016rxo}. 
However, in recent years, it has been argued that more general stable, self-gravitating scalar field objects could exist in nature, and this may affect the previous conclusions. 

An interesting example of such configurations are the $\ell$-boson stars we have presented in previous work~\cite{Alcubierre:2018ahf}.
Based on similar ideas used previously in the context of gravitational collapse~\cite{Olabarrieta:2007di}, $\ell$-boson stars incorporate some effects of the angular momentum into the 
scalar fields while maintaining the spherical symmetry of the spacetime, which results in a relatively simple model for their description. 
In the particular case where $\ell=0$ the standard boson stars by Kaup~\cite{Kaup68} and Ruffini and Bonazzola~\cite{Ruffini69} are recovered. 
However, in general, in addition to the parameters that characterize the standard $\ell=0$ solutions, there is an ``angular momentum number'' $\ell$ 
that provides a model with a richer structure that could potentially be relevant for the description of dark matter halos
and compact objects. In particular, and as we further explore in this article, boson stars with $\ell > 0$ can be more compact than standard ones.
It turns out that the maximum mass of these objects increases greatly with $\ell$, giving masses that are orders of magnitude larger than for the $\ell=0$ case. 
Even if these configurations are also larger in size than the standard ones, the growth in mass is faster than the growth in size in such a way that the compactness increases.

The stability of $\ell$-boson stars under spherical perturbations has first been studied in~\cite{Alcubierre:2019qnh} by performing numerical evolutions of the Einstein-Klein-Gordon
equations in spherical symmetry, and later also in~\cite{Alcubierre:2021mvs} based on a more formal study of the linearized system. 
These analyses have revealed that $\ell$-boson stars show stability characteristics that are qualitatively similar to those of the $\ell=0$ case,
where for each value of $\ell$ there exist a stable and an unstable branch with the transition point given by the solution of maximum total mass.
For other studies addressing the stability of $\ell$-boson stars which are based on full nonlinear numerical evolutions without symmetries 
see~\cite{Jaramillo:2020rsv,Sanchis-Gual:2021edp} (see also~\cite{Guzman:2019gqc} for a study of the Newtonian regime in axial symmetry).
In particular, in~\cite{Sanchis-Gual:2021edp} it was shown 
that $\ell$-boson stars assume a privileged role among other
stationary solutions of the multi-field, multi-frequency scalar 
field scenario as far as their stability is concerned.

In the present work we perform an exhaustive
exploration of the $\ell$-boson stars' parameter space, focusing in particular on solutions with very large values of $\ell$, including the $\ell\rightarrow\infty$ limit. 
Our analysis covers the stars' morphology, anisotropy and compactness, the characteristics of the circular orbits (including the
null ones, also known as light rings), as well as the scaling properties of the fields and relevant physical quantities with respect to $\ell$.
We start in section~\ref{basics} with a brief review of $\ell$-boson
stars, presenting the main equations and properties, including the
definitions of density, pressure, anisotropy and compactness, and present the
equations for geodesic motion, particularly those describing circular
causal geodesics. Next, in section~\ref{Sec:extreme},
we present our solutions, analyzing in each case the
  role played by the angular momentum parameter $\ell$ on various of
  their properties, and paying particular attention to the large $\ell$ regime.
We accomplish this by numerically obtaining and analyzing hundreds of solutions.
The observed scaling properties of the fields for large $\ell$
motivate the in-depth study of section~\ref{s:scaling}, where we obtain
effective equations which describe the asymptotic behavior of the
fields in the limit $\ell\rightarrow\infty$. Conclusions
and an overview of our results are given in section~\ref{Sec:Conclusions}.
Technical details and tables summarizing our notation and numerical data are
included in appendices~\ref{tables}--\ref{table_data}.

Throughout this work we use the signature convention $(-,+,+,+)$ for
the spacetime metric and Planck units such that $G=c=\hbar=1$.
We present our results
in a form that is independent of the scalar field mass $\mu$. The rescaling rules in $\mu$
are summarized in table~\ref{rescaling} of appendix~\ref{tables}.

\section{\texorpdfstring{$\ell$}{l}-boson stars}
\label{basics}

In this section we summarize the relevant equations that describe 
$\ell$-boson stars, as well as some of their most significant properties. 
Additional information can be found in our previous 
works~\cite{Alcubierre:2018ahf,Alcubierre:2019qnh,Alcubierre:2021mvs}; see also 
the review 
articles~\cite{Jetzer92,Liddle:1992fmk,Mielke:1997re,Liebling_2017,
visinelli2021boson} for the standard $\ell=0$ boson stars. 
$\ell$-Boson stars are self-gravitating objects that consist of an odd
number $N = 2\ell+1$ of complex scalar fields $\Phi_{\ell m}$, $m = -\ell,\ldots,\ell$ of equal mass $\mu$ and the same radial profile. The dynamics of these fields is described by the following Lagrangian
\begin{equation}\label{eq:lagrangian_sf}
\mathcal{L}=\frac{R}{16\pi}-\frac{1}{2}\sum_{m=-\ell}^\ell
\left(\nabla_\mu\Phi_{\ell m}\nabla^\mu \Phi^*_{\ell m}+
\mu^2{|\Phi_{\ell m}|}^2 \right) ,
\end{equation}
where $R$ is the Ricci scalar and the scalar fields have the form:
\begin{equation}
\Phi_{\ell m}(t,r,\vartheta,\varphi) = e^{i\omega t}\psi_\ell(r) Y^{\ell m}(\vartheta,\varphi),
\end{equation}
with $\omega$ a real frequency and $\psi_\ell$ a real-valued radial
function which is independent of $m$. As usual, $Y^{\ell m}$ denote the standard spherical harmonics with angular momentum numbers $\ell$ and $m$.
By applying the addition theorem for spherical harmonics  one can see that in the absence of self-interactions, the total stress energy-momentum tensor 
\begin{equation}\label{eq.EM}
T_{\mu\nu} = \frac{1}{2}\sum_{m=-\ell}^\ell\left[\nabla_\mu\Phi_{\ell m}^*\nabla_\nu\Phi_{\ell m} + 
\nabla_\mu\Phi_{\ell m}\nabla_\nu\Phi_{\ell m}^*
- g_{\mu\nu}\left( \nabla_\alpha\Phi_{\ell m}^*\nabla^\alpha\Phi_{\ell m} + \mu^2\Phi_{\ell m}^*\Phi_{\ell m} \right)\right] 
\end{equation}
is spherically symmetric, even if $\ell>0$ ($N>1$) and the individual fields have
angular momentum.

The spacetime metric is parameterized according to 
\begin{equation}
ds^2 = -\alpha^2(r) dt^2 + \gamma^2(r) dr^2 + r^2 d\Omega^2  ,\qquad
\gamma^2(r) := \frac{1}{1 - \frac{2M(r)}{r}}  ,
\label{Eq:Metric}
\end{equation}
where $\alpha$ and $M$ denote the lapse and the Misner-Sharp mass
functions, respectively, $r$ is the areal radius and $d\Omega^2$ is
the standard metric on the unit two-sphere.  The field equations are
obtained from the Einstein-Klein-Gordon system and take the
form~\cite{Alcubierre:2018ahf}:
\begin{subequations}\label{Eq:bosonstars}
\begin{eqnarray}
&& M' = \frac{\kappa_\ell r^2}{2}
\left[ \frac{\psi_\ell'^2}{\gamma^2}
 + \left(\mu^2 + \frac{\omega^2}{\alpha^2} + \frac{\ell(\ell+1)}{r^2} \right)\psi_\ell^2 \right]
  = 4\pi r^2\rho ,
\label{Eq:bosonstars.2} \\
&& \frac{(\alpha\gamma)'}{\alpha\gamma^3} 
 = \kappa_\ell r\left[ \frac{\psi_\ell'^2}{\gamma^2} + \frac{\omega^2}{\alpha^2}\psi_\ell^2 \right]
 = 4\pi r(\rho + p_r) ,
\label{Eq:bosonstars.3} \\
&& \frac{1}{r^2\alpha\gamma}\left( \frac{r^2\alpha}{\gamma}\psi_\ell' \right)' 
 = \left(\mu^2 - \frac{\omega^2}{\alpha^2} + \frac{\ell(\ell+1)}{r^2} \right)\psi_\ell ,
\label{Eq:bosonstars.1}
\end{eqnarray}
\end{subequations}
whith $\kappa_\ell := 2\ell + 1$, and where we have introduced the
energy density, radial pressure and tangential pressure defined as:
\begin{subequations}
  \begin{align}
  \rho:= -{T^t}_t = & \frac{\kappa_\ell}{8\pi} \left[\;\;\; \frac{{\psi_\ell^\prime}^2}{\gamma^2} +
    \frac{\omega^2}{\alpha^2}\psi_\ell^2 +
    \left(\mu^2+\frac{\ell (\ell+1)}{r^2}\right) \psi_\ell^2 \right] , \\
  p_r := {T^r}_r =& \frac{\kappa_\ell}{8\pi} \left[\;\;\; \frac{{\psi_\ell^\prime}^2}{\gamma^2} +
    \frac{\omega^2}{\alpha^2}\psi_\ell^2 -
    \left(\mu^2+\frac{\ell (\ell+1)}{r^2}\right) \psi_\ell^2 \right] , \label{eq.radialpressure}\\
  p_T := {T^\theta}_\theta  = {T^\varphi}_\varphi=& \frac{\kappa_\ell}{8\pi} \left[ -\frac{{\psi_\ell^\prime}^2}{\gamma^2} +
    \frac{\omega^2}{\alpha^2}\psi_\ell^2 -
     \mu^2 \psi_\ell^2 \right] .\label{eq.transversepressure}
  \end{align}      
\end{subequations}
We denote by $M_T$ the total mass of the object, given by the limit $r\to \infty$ of the function $M(r)=4 \pi
\int_0^r \, \tilde{r}^2\,\rho(\tilde{r})\, d\tilde{r}$. In the case of our numerical solutions, we approximate $M_T$ by evaluating $M(r)$ a the outer boundary of the numerical domain (after ensuring that the mass variation is negligible near that boundary).

Each $\ell$-boson star solution is uniquely determined by a given set
of the parameters $\ell$, $\mu$, $u_0$, and a discrete set of values
$\omega$, with $u_0$ given by $\psi_\ell/r^\ell$ evaluated at
$r=0$\footnote{Note that $\psi_\ell=A r^\ell+\mathcal{O}(r^{\ell+2})$ with constant $A$, such that $\psi_\ell/r^\ell$ is regular at $r=0$. In practice $u_0$ is evaluated either by taking the limit $r\rightarrow0$ or by directly evaluating $u_0=A$ (see also Appendix~\ref{Sec:Methods}).}. Given $\ell$ and $\mu$, $u_0$ is a free parameter (which
reduces to the central scalar field amplitude $\psi_c=\psi(r=0)$ for
$\ell=0$), and the $\omega$'s are the frequency eigenvalues obtained
by demanding that the field vanishes at infinity and that the solution
remains regular at $r=0$.  In this work we only consider the ground
state for which $\psi_\ell$ has no nodes in the open interval
$r\in(0,\infty)$, hence fixing $\omega$ for each $\ell$ and $u_0$.  Finally,
solutions with different $\mu$ are related to each other by a simple rescaling (see
table~\ref{rescaling} in appendix~\ref{tables}).
Consequently, for each $\ell$ it is sufficient to study a one-parameter family of solutions, usually parameterized by $u_0$ or
(equivalently) by $\alpha_0:=\alpha(r=0)$.

\begin{figure}
\includegraphics[width=0.65\textwidth]{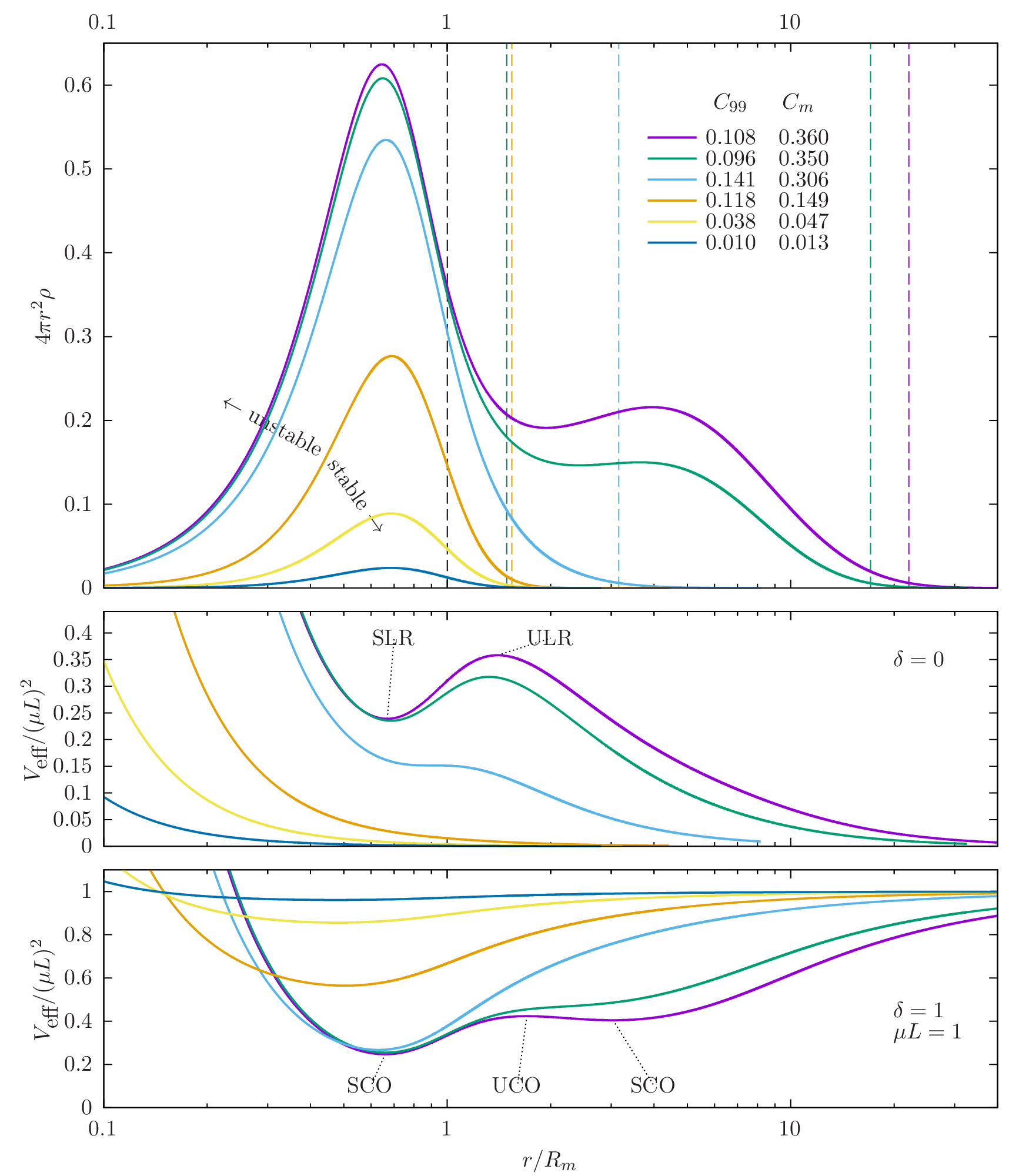}
\caption{{\em Top panel:} 
For various configurations with $\ell=1$, we show the
  rescaled  density profile with respect to the rescaled radial coordinate $r
  \rightarrow r/R_m$. This rescaling provides an easier way to compare
  the different curves between themselves, while also allowing to easily
  locate $R_m$, given by $r/R_m=1$ (black vertical line), as well as $R_{99}$
  (vertical lines with the same color as the
  corresponding solution).
  {\em Middle panel:}
  Effective potentials for circular null geodesics ($\delta=0$) for
  the same solutions as in the top panel, displaying cases without
  light rings, with a pair of (stable and unstable) light rings, and
  the transition solution for which the (degenerate) light rings first
  appear.
  {\em Bottom panel:} Effective potentials for circular timelike
  geodesics ($\delta=1$) with $\mu L=1$ for the same solutions as in the
  other panels. The local extrema correspond to stable (minimum) and
  unstable (maximum) circular orbits.}
 \label{Vrho}
\end{figure}

Since boson stars do not have a well defined boundary, one usually
describes their size by the $R_{99}$ radius, defined as the (areal)
radius of the sphere containing 99\% of the total mass $M_T$.  In
addition, we use two different measures for the star's compactness:
\begin{subequations}\label{eq.compactness}
\begin{equation}
C_{99} := \frac{M_T}{R_{99}} ,
\end{equation}
and
\begin{equation}
C_m := \max\limits_{r > 0} \left\{ \frac{M(r)}{r} \right\} =: \frac{M_m}{R_m} ,
\label{Cmax}
\end{equation}
\end{subequations}
where we also defined $R_m$ as the point $r$ of maximum $M(r)/r$, and
$M_m$ as $M(r=R_m)$.  To help better understand the meaning of these definitions we
highlight their differences in the top panel of figure~\ref{Vrho},
where some density profiles are shown, together with vertical lines
indicating the radii $R_{99}$ and $R_m$ for each star.

As one can appreciate from this figure, some solutions (see for instance the purple and green lines) can be interpreted as having two parts: A very compact ``core'', located
mostly to the left of $r/R_m=1$, plus a less dense ``halo'' to the
right of that point. Note that the halo is much wider than the central
region (a fact that might be unnoticed at a first glance since the
horizontal axis is in logarithmic scale). We clearly see that the
definition $C_{99}$ is a proper indicator of the whole object's
compactness, while the definition $C_m$ is more representative of the
central region's compactness.  However, as we will see later, the sets of definitions $\left[R_{99}, M_T, C_{99}\right]$ and $\left[R_m, M_m,
  C_m\right]$ tend to coincide for larger $\ell$'s.\footnote{See also
figure~\ref{LR_Cmax} for noticeable differences between the two sets
of definitions.}
Although we have found the core-and-halo structure only for configurations lying on the unstable branches, the mentioned differences between these two sets are seen for stable as well as unstable solutions.

The stress tensor of a perfect fluid is isotropic, and pressure is the same in all directions of a fluid star. Even if
common for some materials, 
isotropy is not a natural consequence of the underlying spacetime symmetries, 
and there exist static and spherical configurations 
that exhibit fractional anisotropy, defined as the relative difference between the radial and tangential components of the pressure:
\begin{equation}
 fa = \frac{p_r-p_T}{p_r} .
\end{equation} 
From the right-hand sides of equations~(\ref{eq.radialpressure}) and~(\ref{eq.transversepressure}) one can see that $\ell$-boson stars
(including the standard $\ell=0$ boson stars) are anisotropic.
Furthermore, one might suspect that solutions with higher anisotropy will exist for the cases with non-vanishing angular momentum number, due to the presence of the centrifugal term $\ell(\ell+1)/r^2$ in $p_r$.
We will corroborate this assertion in the next sections.
This is not just a curious fact,
since configurations with larger fractional anisotropy have been identified to be stable up to higher values of the central density~\cite{Gleiser:1988rq},
hence leading to more compact objects~\cite{Dev:2000gt}. 
This enhancement in the allowed compactness within the stable branch is an interesting property 
that is also satisfied for $\ell$-boson stars, as we discuss later.

It will also be helpful to identify some general properties of the motion of test particles propagating in the spacetime associated with the $\ell$-boson stars, and in particular to determine 
whether  the solutions admit innermost stable circular orbits (ISCOs) and/or light rings~\cite{Cardoso:2019rvt}
and, if so, to find their location.
Given the spacetime symmetries we can obtain the geodesics with the help of
conserved quantities using the expression~\cite{Wald84}
\begin{equation}
  \left( \frac{d r}{d \lambda} \right)^2 = \frac{E^2}{\alpha^2 \gamma^2} - \frac{1}{\gamma^2}
  \left( \delta + \frac{L^2}{r^2} \right),
\label{Eq:GeodesicMotion}
\end{equation}
where $E$ and $L$ are constants of motion (associated with the particle's
energy and total angular momentum), and $\delta=0$ for null geodesics, while $\delta=1$ for timelike
geodesics. 
It is convenient to introduce the effective potential
\begin{equation}
  V_{\rm eff}(r):=\alpha^2 \left( \delta + \frac{L^2}{r^2} \right) , \label{eq:veff}
\end{equation}
leading to an equation of motion that resembles a point particle moving in a one-dimensional potential.\footnote{Defining $x:=\int_0^r\alpha(r)\gamma(r)dr$ we can rewrite equation~(\ref{Eq:GeodesicMotion}) as
  $\left( \frac{dx}{d\tau}
  \right)^2=E^2-U_{\rm eff}(x)$, where
$U_{\rm eff}(x):=V_{\rm eff}[r(x)]$. Hence, in analogy with Classical
Mechanics we can infer that the orbits are restricted to the 
regions where $U_{\rm eff} \leq E^2$, with the equality being satisfied at the
turning points. Circular orbits are obtained where $E^2$ equals an extremum of
$U_{\rm eff}$, and their stability depends on whether the extremum is a maximum
or a minimum. Given that the transformation $x=x(r)$ is monotonic, the same conditions
are satisfied for $V_{\rm eff}(r)$.}
Then, orbiting particles are restricted to the regions where $V_{\rm
  eff}(r)<E^2$. Circular orbits can be obtained when $E^2$ equals a local extremum
of $V_{\rm eff}$, and those orbits are stable (unstable) if said extremum is a
minimum (maximum).

In the null case, the condition for circular orbits is
\begin{equation}\label{Eq:LRcondition}
\alpha - r \, \alpha^\prime = 0 , 
\end{equation}
where the sign of the second derivative of the lapse function evaluated at the light ring radius determines the stability of the orbit: it is stable if $\alpha''$ is negative and unstable 
otherwise~\cite{Barranco:2021auj}.
In the timelike case the energy and total angular momentum per unit rest mass of a particle in circular motion at radius $r$ must satisfy  
\begin{equation}
 E= \sqrt{\frac{\alpha^3}{\alpha-r\alpha'}}, \quad L = \sqrt{\frac{r^3\alpha'}{\alpha-r\alpha'}} .
\end{equation}
These orbits are stable wherever $L(r)$
grows with $r$, whereas they are unstable otherwise~\cite{Barranco:2021auj}.
For regular configurations light rings can appear only in pairs, one of
them being stable and the other unstable. Note, however, that not all
  stars admit light rings.
On the other hand, there always exist stable circular 
orbits of massive particles. 
In particular, the existence of stable orbits is
  guaranteed both at large distances and close enough to the center. However,
regions of instability may exist too, being delimited by 
innermost stable circular orbits (ISCOs) and  outermost stable circular
orbits (OSCOs).
In a similar way, the light ring pairs delimit a region where $\alpha - r \, \alpha^\prime$ is negative and circular orbits are not allowed at all.\footnote{We note that in all the solutions 
we have found $\alpha'(r) > 0$ for $r > 0$, such that the lapse is monotonously increasing.}
We will now give more explicit details about these assertions.

The central panel of figure~\ref{Vrho} illustrates distinct cases regarding the existence of light rings, as determined by equation (\ref{eq:veff}),
all with $\ell=1$: (i)~Potentials without local extrema (besides at $r=0$). These solutions cannot have light rings.
(ii)~Potentials with a local minimum at some $r=r_{\rm in}$ and with a local maximum at
some other $r=r_{\rm out}$, such that $r_{\rm in}<r_{\rm out}$. These
solutions have a pair of light rings, a stable one at $r_{\rm in}$ and an
unstable one at $r_{\rm out}$. (iii)~The transition case, in which the potential
have an inflection point, giving rise to degenerate light ring solutions with
$r_{\rm in}=r_{\rm out}$. 
Note that cases (ii) and (iii) 
only occur for unstable spacetimes~\cite{Alcubierre:2019qnh,Alcubierre:2021mvs}.
In a similar way, in the bottom panel of this figure we illustrate different cases regarding the existence of unstable circular orbits of massive particles with $\mu L=1$.

Finally, we give an expression for the test particle's speed moving on a circular orbit (more precisely, the magnitude of its three-velocity as measure by a static observer located at the corresponding radius):
\begin{equation}
  v(r) := r \frac{d\phi}{dt} = \sqrt{ \frac{r \alpha^\prime(r)}{\alpha(r)} },
  \label{eqRC}
\end{equation}
which will be used in the next section to show some rotation curves.

\section{Extreme \texorpdfstring{$\ell$}{l}-boson stars}
\label{Sec:extreme}

In this section we present and analyze our results. 
For all integer $\ell$ from $0$ to $15$, and for $\ell=20$, $25$,
$50$, $75$, $100$, $200$, $400$ and~$1600$, we constructed solutions, tens of them in some cases,  
that correspond to
different values of the central parameter $u_0$.
The parameters and main
properties of some of the most relevant solutions that we have obtained are displayed in
table~\ref{data} of appendix~\ref{table_data}, which also includes a reference
to the figures in which they  are used.
In addition, in the next
section we obtain general expressions that are applicable for the limiting case in which $\ell\rightarrow\infty$.

\begin{figure}
\includegraphics[width=0.65\textwidth]{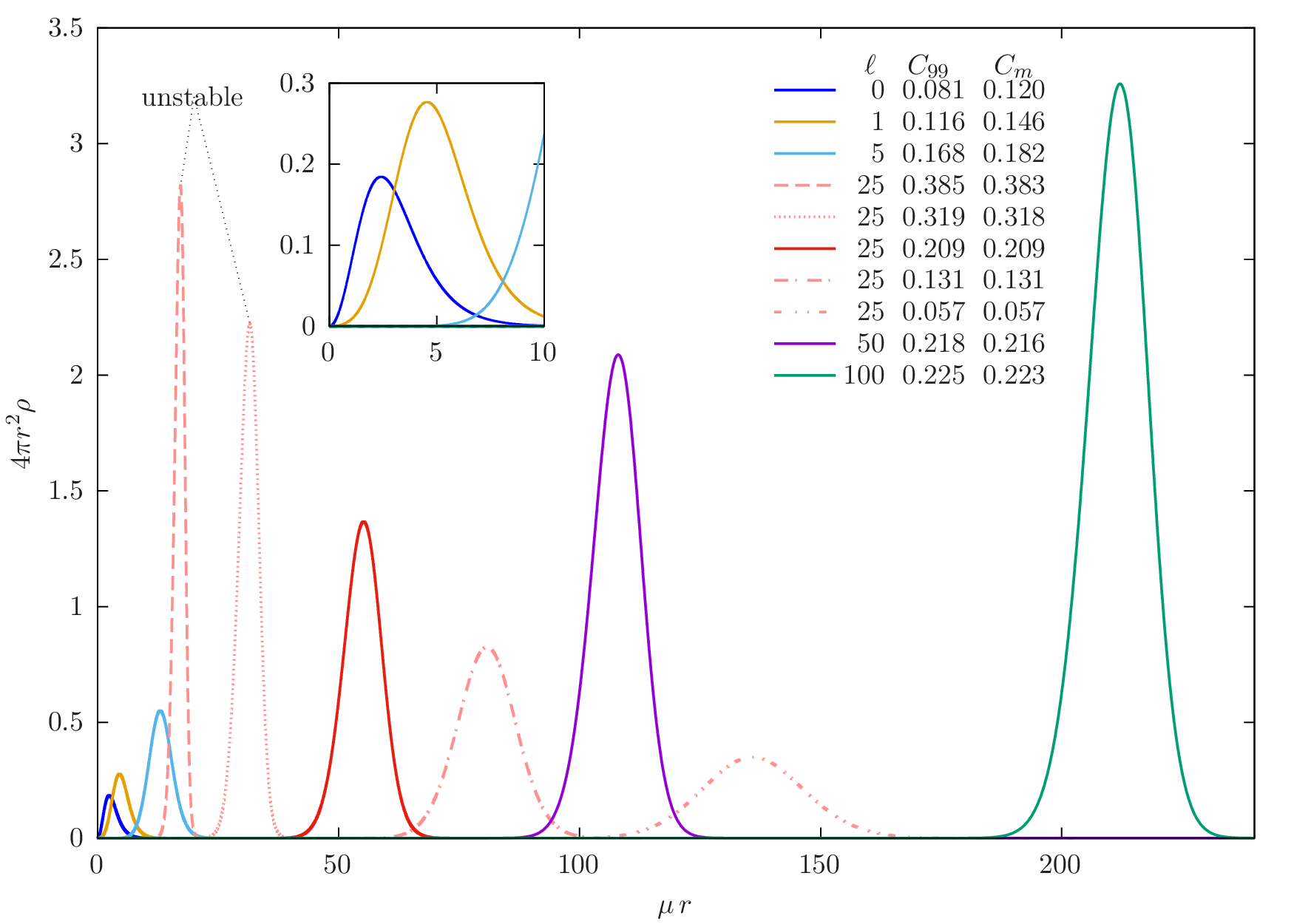}
\caption{Rescaled  density profiles, $4\pi r^2\rho(r)$,  of solutions with maximum $M_T$ for various
  values of $\ell$ (solid lines), and of varying compactness 
  for fixed $\ell=25$, both in the stable and unstable region (red lines).
  The inner panel shows a zoom into the small $r$ region
  for a better reference of the cases with $\ell=0$ and $1$.
}
\label{density} 
\end{figure}

We present some of our solutions in figure~\ref{density}, where we show the
rescaled density profiles (defined as
$\varrho=4\pi r^2\rho$ such that $M=\int \varrho dr$) associated with some of our configurations.\footnote{Throughout 
this section we alternate between showing results in terms of $\rho$ and $\varrho$,
depending on what we find more illustrative.}
Since one needs some criterion in order to compare
solutions through different values of $\ell$, in this case we chose to display
configurations that, for each $\ell$, have the maximum total mass, which are
also the most compact stable solutions.
This is a criterion we will adopt in most of this work.
In the same figure we also show some solutions for given $\ell$ ($=25$) and varying compactness, the more compact ones
being unstable.
The solutions clearly exhibit a shell-like morphology, at least for $\ell > 1$.
For bigger $\ell$ the stars are larger both in size and in total
mass. We will see that
the compactness also increases with $\ell$. In contrast, if one considers stars
with fixed $\ell$ and increasing size, 
the compactness decreases. We also note that, as is the case for the traditional $\ell=0$
boson stars, the most compact solutions belong to the unstable branch.

\begin{figure}
  \includegraphics[width=0.42\textwidth]{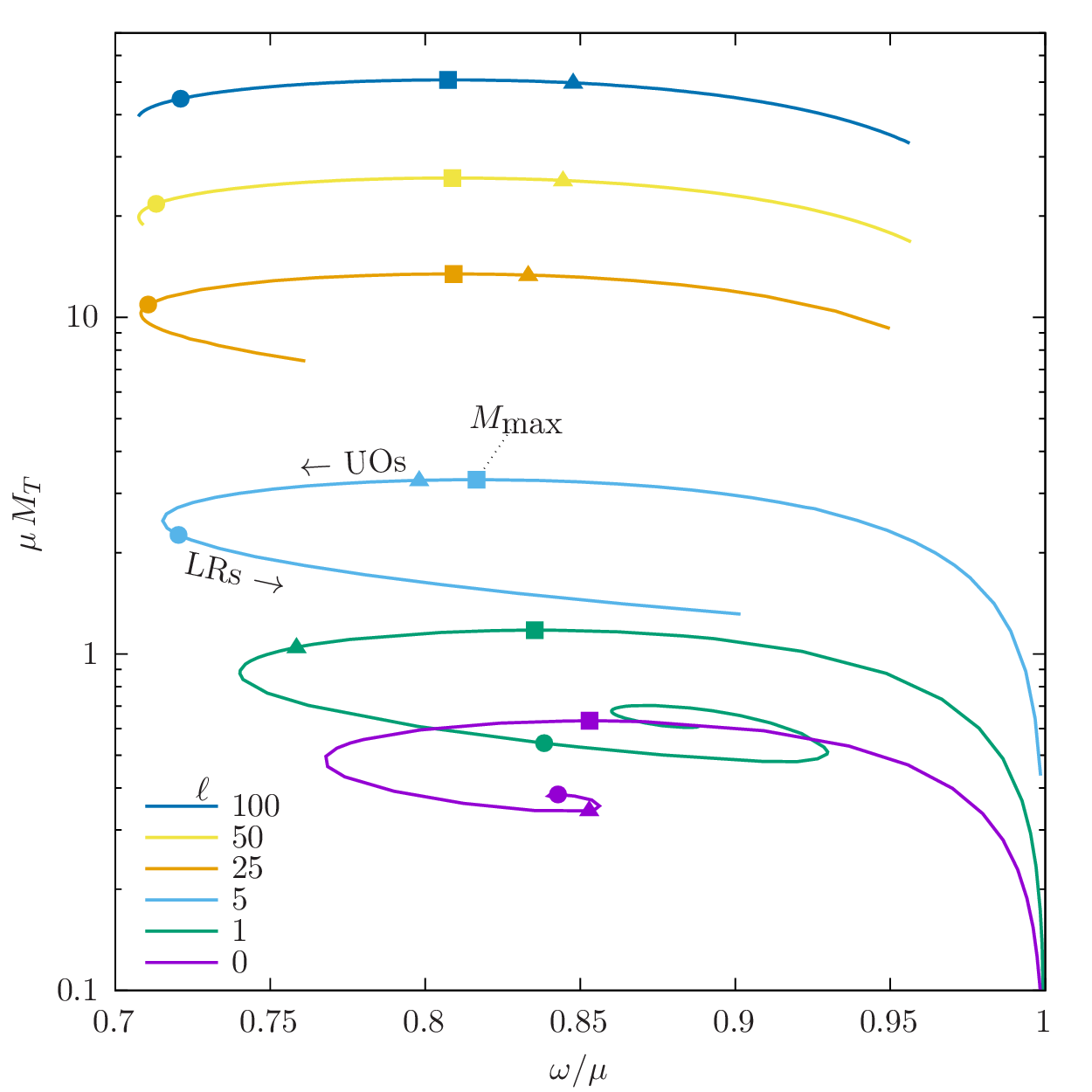}
  \includegraphics[width=0.42\textwidth]{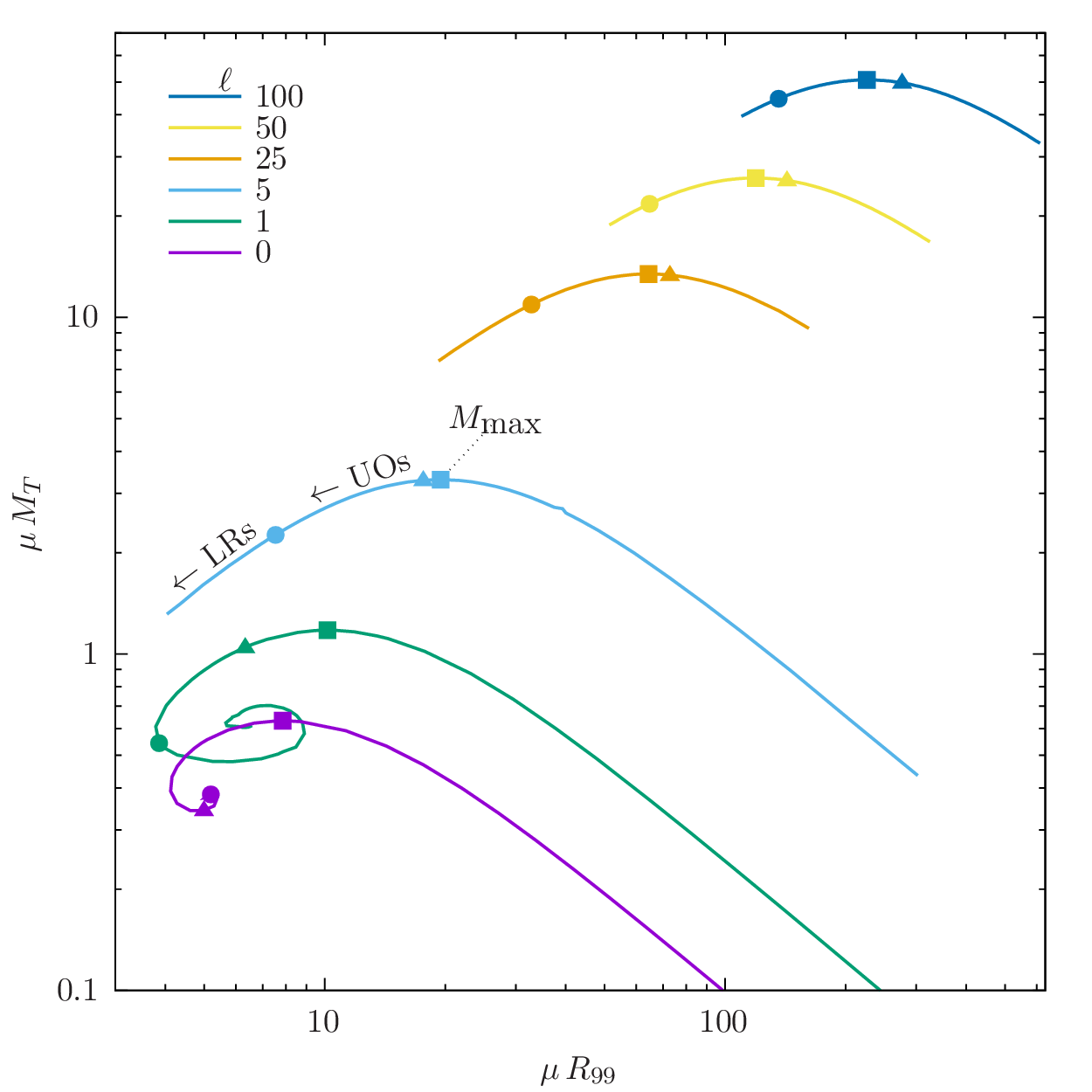}
\caption{$M_T$ vs. $\omega$ and vs. $R_{99}$ for $\ell=0$, $1$, $5$, $25$, $50$
  and $100$. Each point on these curves corresponds to a different solution, including
  for instance those shown in figure~\ref{density}.
  The
  squares denote the maximum of the total mass, which separates the stable and
  unstable regions. The circles denote the first appearance of light rings,
  while the triangles denote the first appearance of an ISCO-OSCO
  pair and, hence,  the existence of unstable orbits (UOs).
}
 \label{Mw_and_Mr99}
\end{figure}

Figure~\ref{Mw_and_Mr99} shows the dependence of the total mass
 on the frequency  and on the $R_{99}$ radius for $\ell=0$, $1$, $5$, $25$,
$50$ and $100$. For each $\ell$ we indicate the maximum of $M_T$
(squares), which we denote $M_{\rm max}$, and the first appearance of a light rings
pair (circles)  and of an ISCO-OSCO pair (triangles).
We have seen in previous
works~\cite{Alcubierre:2019qnh,Alcubierre:2021mvs} that the state of maximum mass marks the transition from the stable
solutions (to the right in these figures) to the unstable ones (to the left) for $\ell$ in the interval from 0 to 5. 
We also corroborated in the present work that this fact is still true for larger values of $\ell$. 

In the following subsections we analyze various properties of these
solutions, including their compactness, anisotropy and causal circular orbits.

\subsection{Compactness}

In this section we explore the compactness of our solutions using the definitions of equations~(\ref{eq.compactness}).
As can be seen from figure~\ref{Mw_and_Mr99}, larger values of $\ell$ lead to
solutions with higher total mass $M_T$. On the other hand,
considering for instance the solutions of maximum mass, the radius also increases with $\ell$, as can be inferred from that same figure and figure~\ref{density}.
However, the increase in mass tends to ``win'' over the
increase in radius in such a way that their ratio, the compactness, increases
with $\ell$. Note that said solutions are the most compact stable ones for each $\ell$.

\begin{figure}
\includegraphics[width=0.3\textwidth]{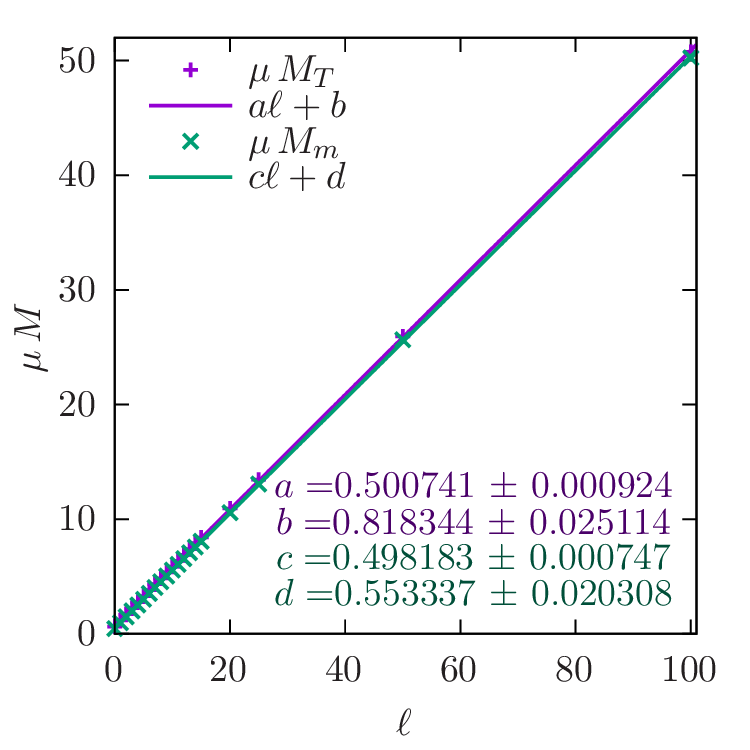}
\includegraphics[width=0.3\textwidth]{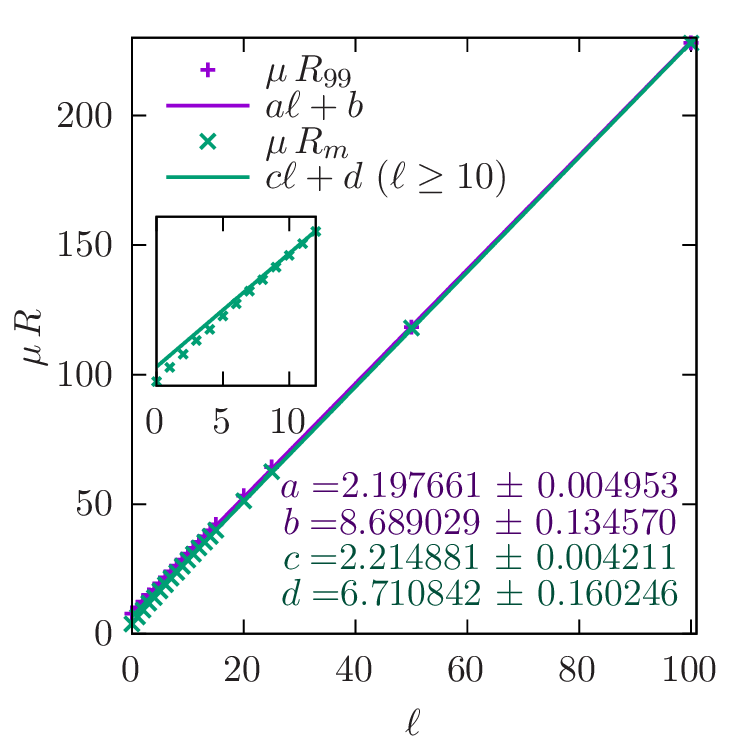}
\includegraphics[width=0.3\textwidth]{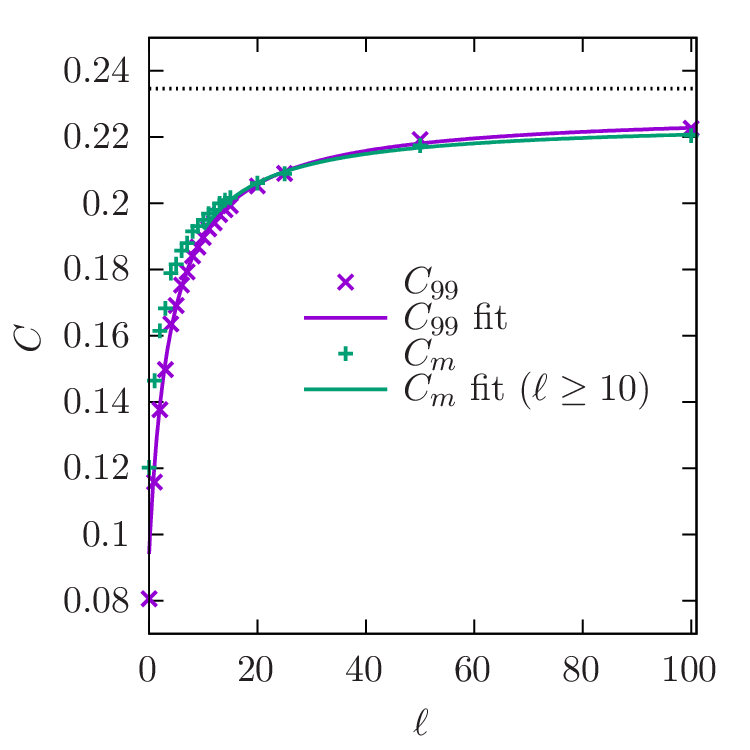}
\caption{We show the dependence on $\ell$ of the stars' mass, radius and
  compactness. All quantities shown here correspond to the solution of maximum
  $M_T$ for each value of $\ell$.
  {\em Left panel:} $M_{T}$ and $M_m$, together with
  the corresponding linear fits.
  {\em Center panel:} $R_{99}$ and $R_m$, together with their linear fit. In the
  case of $R_m$ we do the linear fit only to the points with $\ell\geq10$,
  which is the
  region  where we actually see a linear dependence.
  {\em Right panel:} Compactness $C_m$ and $C_{99}$, and, in each case, the
  compactness calculated from the fits of the previous panels. We also
  indicate the asymptotic value  as a dotted line (see
  section~\ref{s:scaling}).
}
 \label{CmaxC99}
\end{figure}

After inspection of our solutions we note that 
the two mass definitions $M_{T}$ and $M_m$ from equations~(\ref{eq.compactness}), as well as their associated radii $R_{99}$ and $R_m$, seem to both show a
linear relation with $\ell$, at least at large $\ell$ ($\ell \gtrsim 10$). 
This can be seen in the first two panels of figure~\ref{CmaxC99}.
Once again, in order to compare configuration with different $\ell$'s between each other, we have chosen those solutions with maximum total mass $M_{\textrm{max}}$ for each $\ell$.

The apparent linear dependence in $\ell$ suggests that simple expressions can be obtained by performing
linear fits. We show the results of said fits in the figure (continuous
lines), together with the fit coefficients ($a$ to $d$) and their respective errors.
Keeping only two significant figures and omitting the errors we can write:
\begin{subequations}
\begin{eqnarray}
  \mu M_T    &\approx& 0.50\,\ell + 0.82, \\
  \mu M_m    &\approx& 0.50\,\ell + 0.55, \\
  \mu R_{99} &\approx& 2.2\, \ell + 8.7, \\
  \mu R_m    &\approx& 2.2\, \ell + 6.7.
\end{eqnarray}
\end{subequations}
From here, expressions for our two definitions of compactness can be found by taking the quotient of each $M$ vs. $R$ pair.
Said quotients, i.e. $C_{99}$ and
$C_{m}$,  are shown in  the last panel of figure~\ref{CmaxC99}.
The point values shown in that panel are obtained by taking individually the quotient
of the corresponding data pairs that appear in the first panels, while the continuous
line represent the quotient of the linear fit's expressions.

The almost linear relations shown in the first two panels of
figure~\ref{CmaxC99} suggest that solutions might have simple rescaling
properties with $\ell$, at least at large enough $\ell$.
A more detailed analysis of such scaling properties will be given
 in section~\ref{s:scaling}, where we will see that an asymptotic value can be obtained for the
compactness at large $\ell$.
That value is indicated in the right panel of figure~\ref{CmaxC99}
 as a dotted line.
Note that initially the compactness  increases rapidly with $\ell$, and
continues to rise monotonically, remaining close to and below the asymptotic
value derived in section~\ref{s:scaling}.

\subsection{Anisotropy}
\label{s:Anisotropy}

\begin{figure}
  \includegraphics[width=0.37\textwidth]{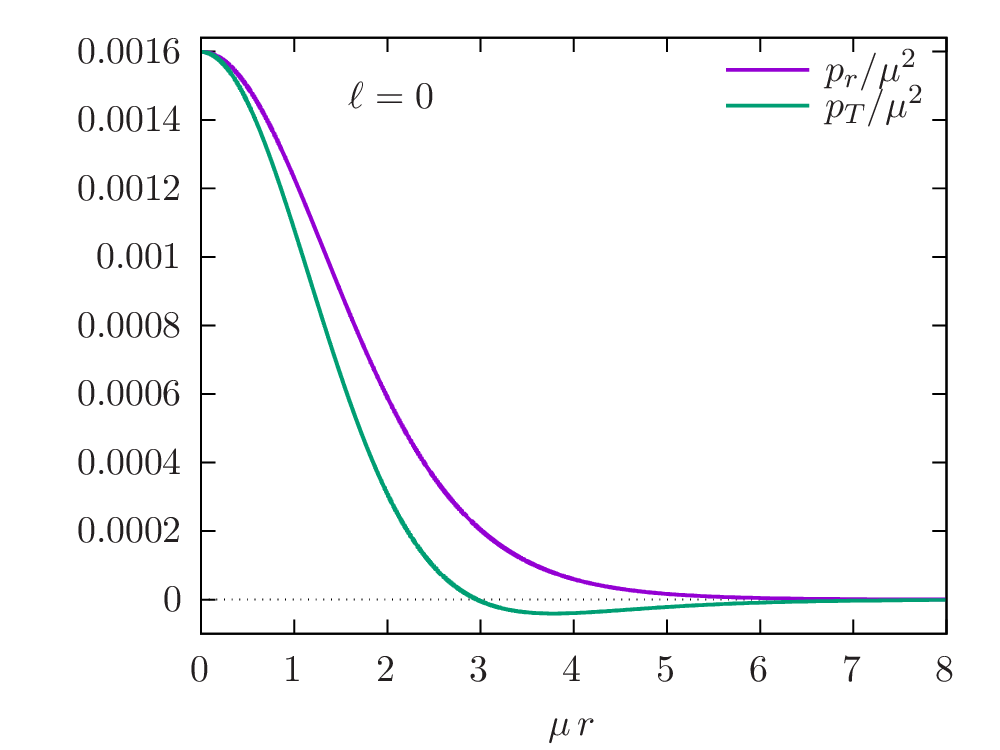}
  \includegraphics[width=0.37\textwidth]{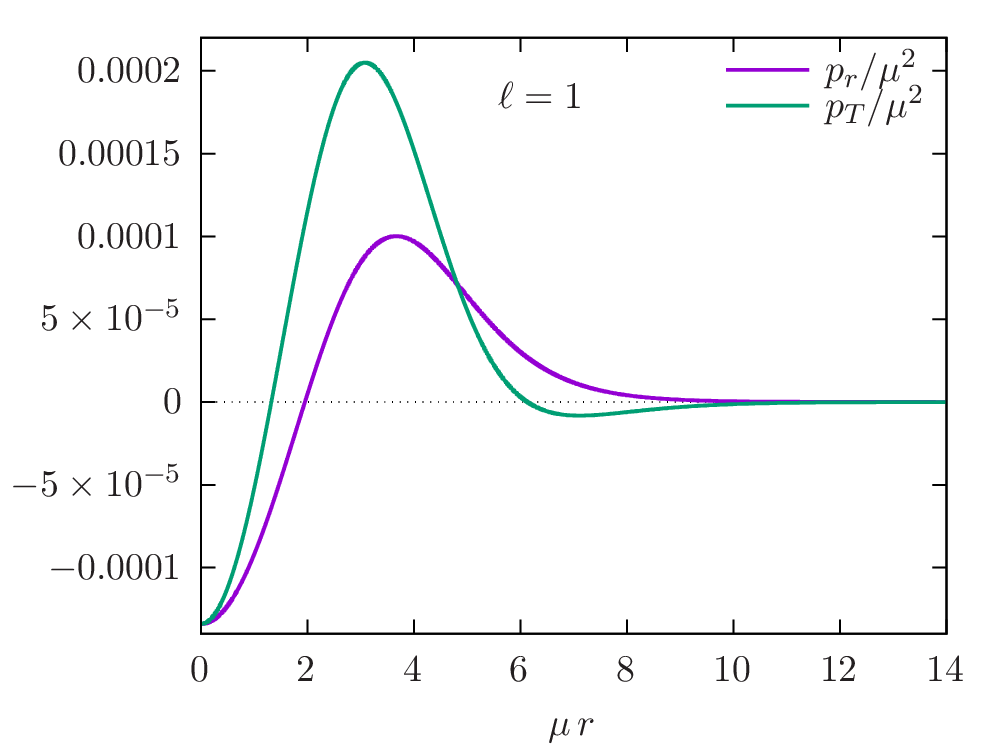}
  \includegraphics[width=0.37\textwidth]{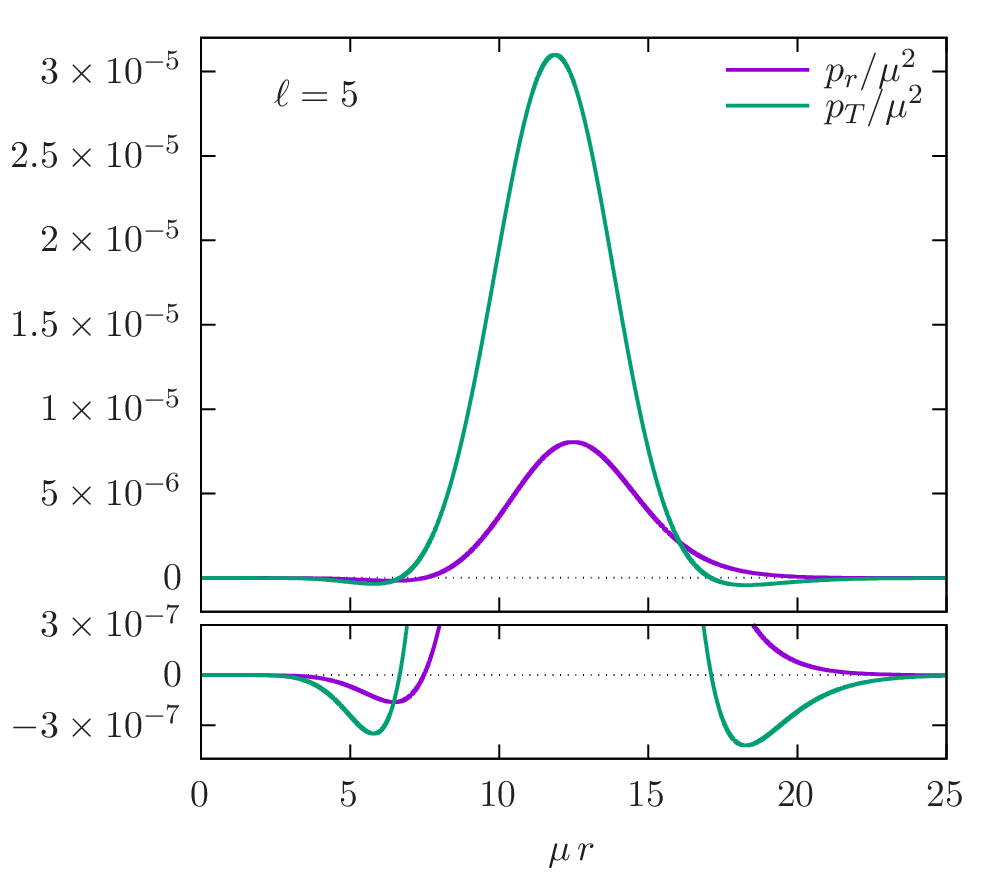}
  \includegraphics[width=0.37\textwidth]{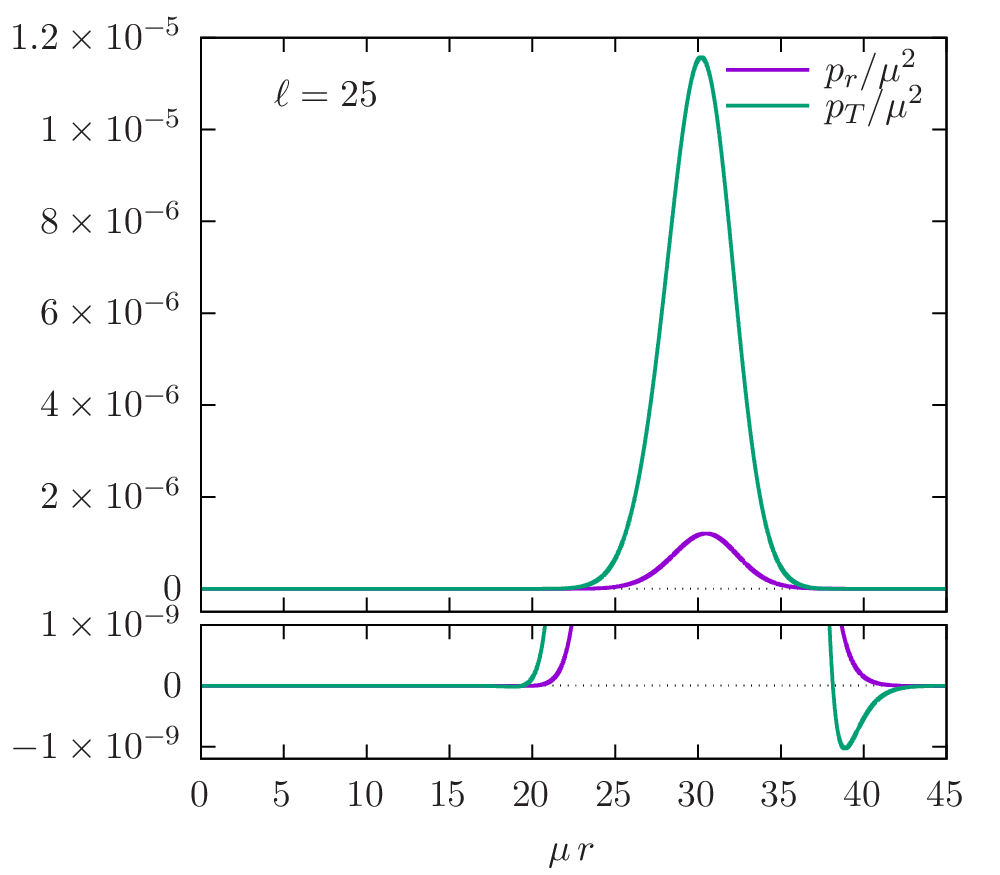}
\caption{Radial and tangential pressures vs. radial coordinate  for the solutions of
  maximum $M_T$ in the cases of  $\ell=0$, 1, 5 and 25. We see that the tangential pressure becomes
  larger and larger relative to the radial pressure when $\ell$ increases.}
 \label{pressures}
\end{figure}

 We move now to the description of the stars' anisotropy. In figure~\ref{pressures} we show the
pressure profiles for $\ell=0 $, $1$, $5$ and $25$, in all cases for the
solution of maximum mass $M_{\rm max}$.
Notice how different the profiles are for $\ell=0$, $\ell=1$, and
$\ell>1$. The typical $\ell=0$  ``solid-sphere'' star has $p_r>p_T$, while for
larger $\ell$'s the ``shell-like'' stars have mostly $p_r < p_T$, with this
difference becoming more pronounced the higher the value of $\ell$.
This behavior seems intuitively natural given the stars' morphology. As $\ell$
increases, the "shells" become larger, as well as thinner relative to their 
radius, in
such a way that the tangential pressure has to become larger relative to the
radial one in order to support the configuration.  

\begin{figure}
  \includegraphics[width=0.95\textwidth]{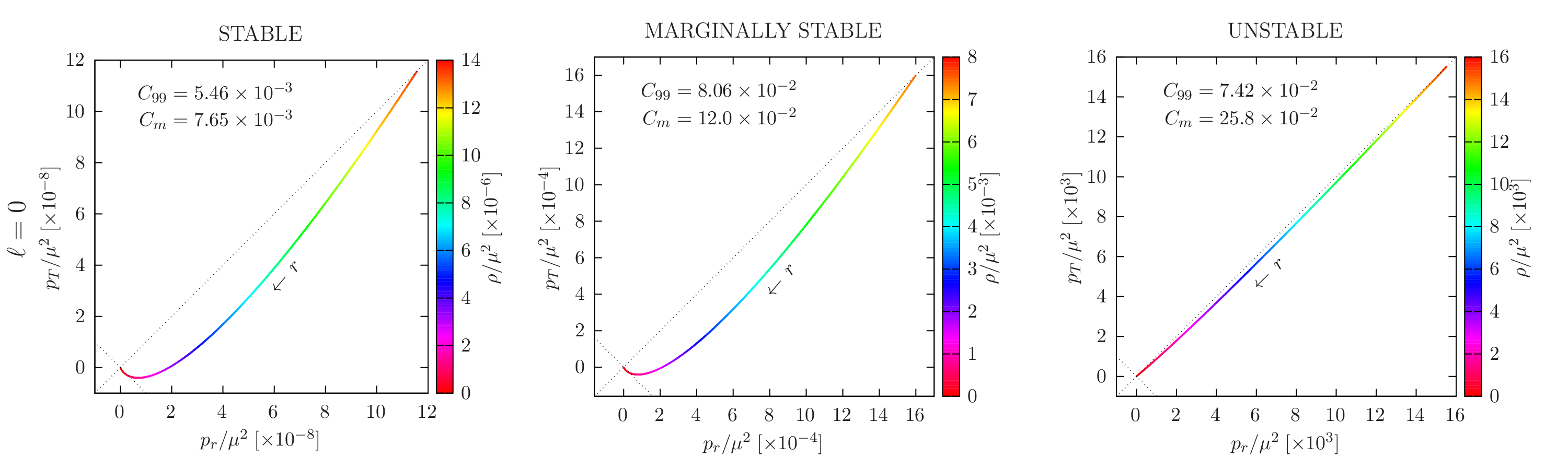}
  \includegraphics[width=0.95\textwidth]{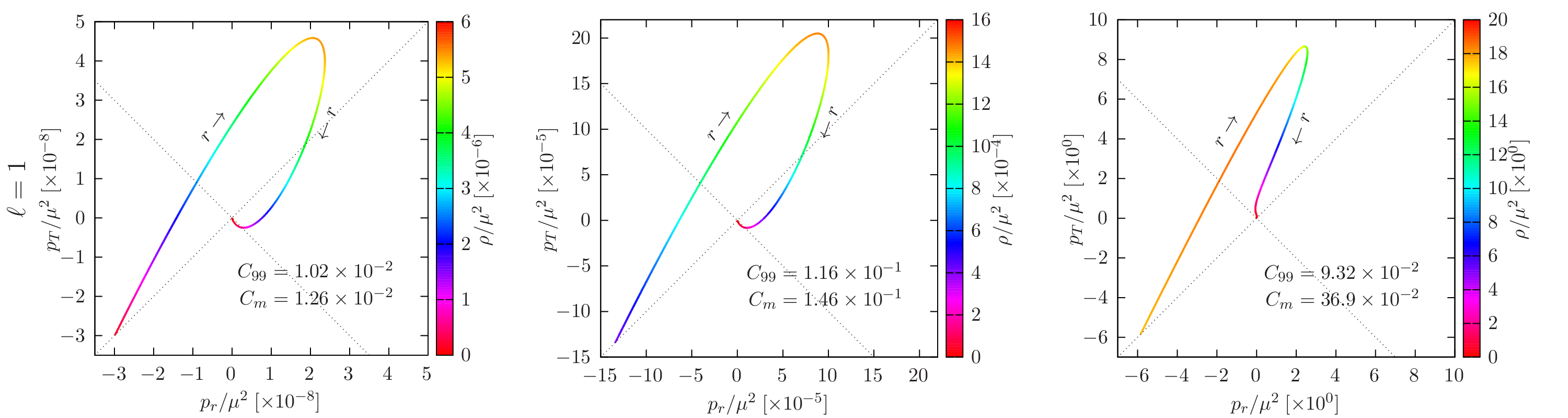}
  \includegraphics[width=0.95\textwidth]{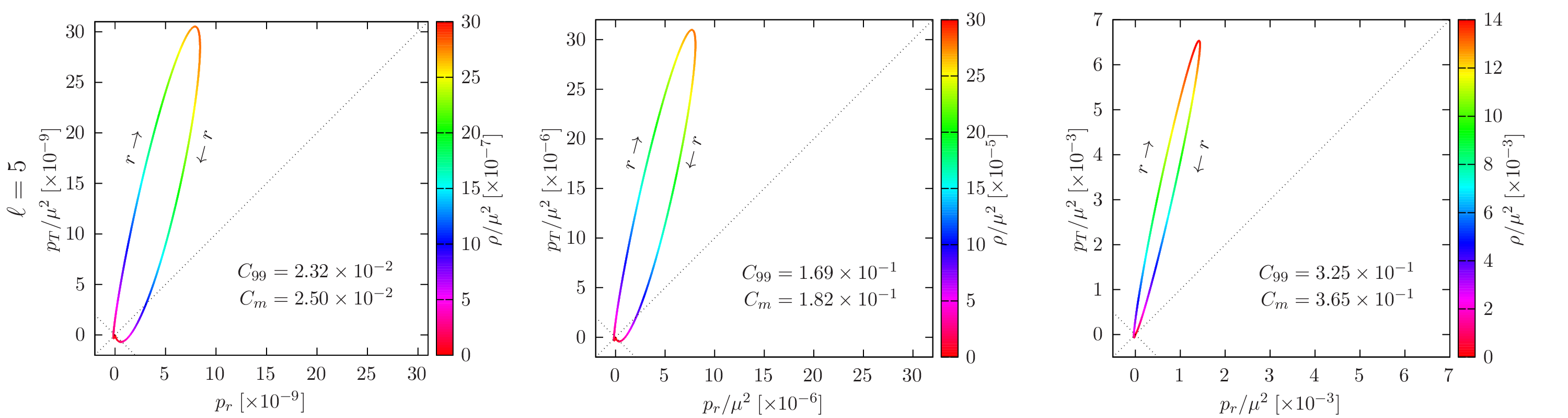}
  \includegraphics[width=0.95\textwidth]{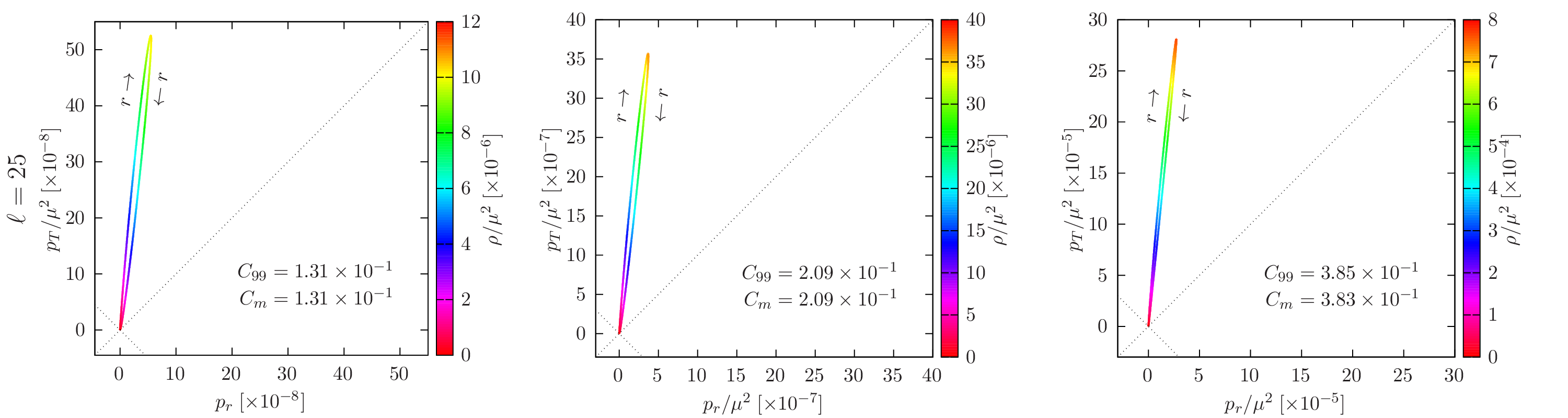}
\caption{Parametric plots of  $[p_r,p_T](r)$. Each row corresponds to a value
  of $\ell$, while each column corresponds to a stability type. These figures
  are particularly well suited for
  analyzing the pressure anisotropies, measured by the deviation from the
  identity $p_r=p_T$. The value of the density $\rho$ along the curves is
  indicated with a color map. The direction of growing $r$
  is also indicated. To help guide the eye, the curves $p_r = \pm p_T$ are
  shown as dotted lines, and the same scale was used in 
  each pair of axes. }
 \label{prpt}
\end{figure}

In figure~\ref{prpt} we show parametric plots of
[$p_r$,$p_T$] vs. $r$, in which the larger the deviation from the identity $p_r = p_T$ (shown as a dotted line of unit slope), the larger the anisotropy.
Additionally, we indicate the
density as a color map, as well as the compactness in each case.
The differences at the starting points of these curves, which correspond to
the pressure values at the origin $r=0$, are consistent with the stars' shape as seen in
our previous work: while they are ``empty'' at the center when $\ell>1$, they
have  maximum density there when $\ell=0$, as is a well known property of standard
boson stars. In the intermediate case, $\ell=1$, the density is greater than
zero at the center, but it does not reach its maximum value at that point.
It is also clear from these plots that the anisotropy, as well as the compactness, 
grow with $\ell$, the tangential pressure becoming larger and larger compared to the radial pressure.
In section~\ref{s:scaling} we will see  that the limiting case $\ell\rightarrow\infty$ would
display a vertical line in this type of plot.
On the other hand, we see little differences in anisotropy
when transitioning between stable and unstable solutions for any given value of $\ell$.

\subsection{Geodesic motion}

Given the large compactness that $\ell$-boson stars may achieve, one may wonder whether they admit light rings and/or ISCOs/OSCOs. In fact, it is known that even traditional $\ell=0$ boson stars can have light rings and ISCOs/OSCOs,
although this is true only in the case of solutions located very deep into the unstable region.
\footnote{Note that the situation may change when non-canonical kinetic terms are considered~\cite{Barranco:2021auj}.}
In the remainder of this section we
will analyze the appearance of light rings and ISCOs/OSCOs, paying particular attention to
their relation with compactness and stability of the underlying spacetime solutions.

In figure~\ref{Mw_and_Mr99} we indicated with a circle
the point corresponding to the first, or less compact, solutions containing a pair of light rings. In all the
cases we studied,  such solutions are always in the unstable region, although
they get closer to the stable region as $\ell$ increases.
It is unclear, however, whether light rings may be found in the stable
region for large enough $\ell$, although one would expect that this is
  not the case given that the maximum  
  compactness that a stable $\ell$-boson star is able to achieve, $C\approx0.235$, is far from the expected one for the appearance of light rings, $C=1/3$.
  The results presented in~\cite{Cunha_2017} seem to indicate  that light rings can only exist for
unstable solutions. A recent work~\cite{guo2021light} also presents results
that support that hypothesis. Another matter of astrophysical interest is whether such
unstable solutions have a relatively short or a rather long life-time. However, this question goes beyond the scope of the present article, so we leave it for future work.

Regarding the existence of ISCOs, we also indicated in
figure~\ref{Mw_and_Mr99} the first appearance of an ISCO-OSCO pair
(triangles). We see that for large enough $\ell$ these pairs can also exist in
the case of stable spacetime solutions. In fact, we have found that the
smallest $\ell$ for which stable $\ell$-boson stars with ISCO-OSCO pairs exist
is $\ell=9$.

\begin{figure}
\includegraphics[width=0.44\textwidth]{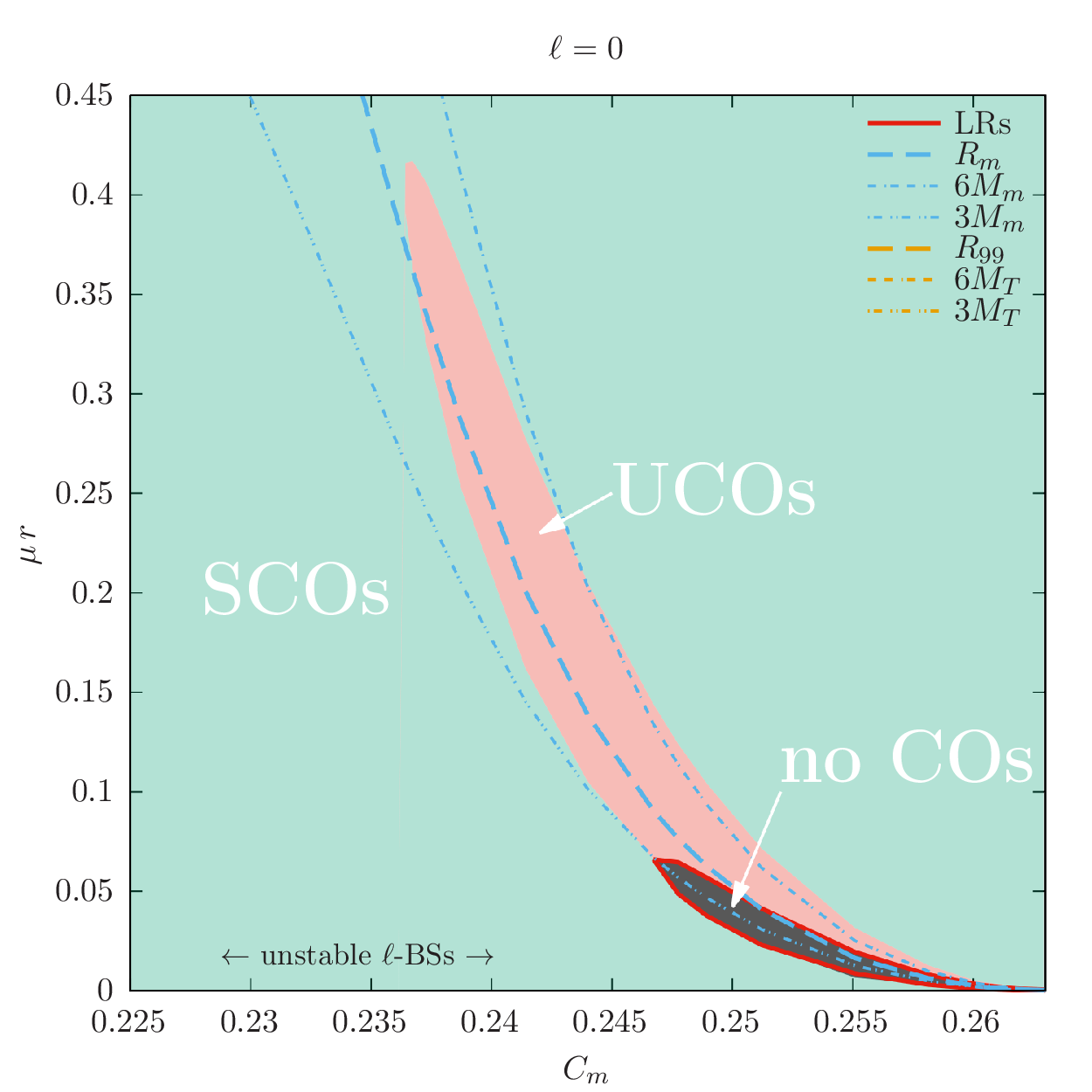}
\includegraphics[width=0.44\textwidth]{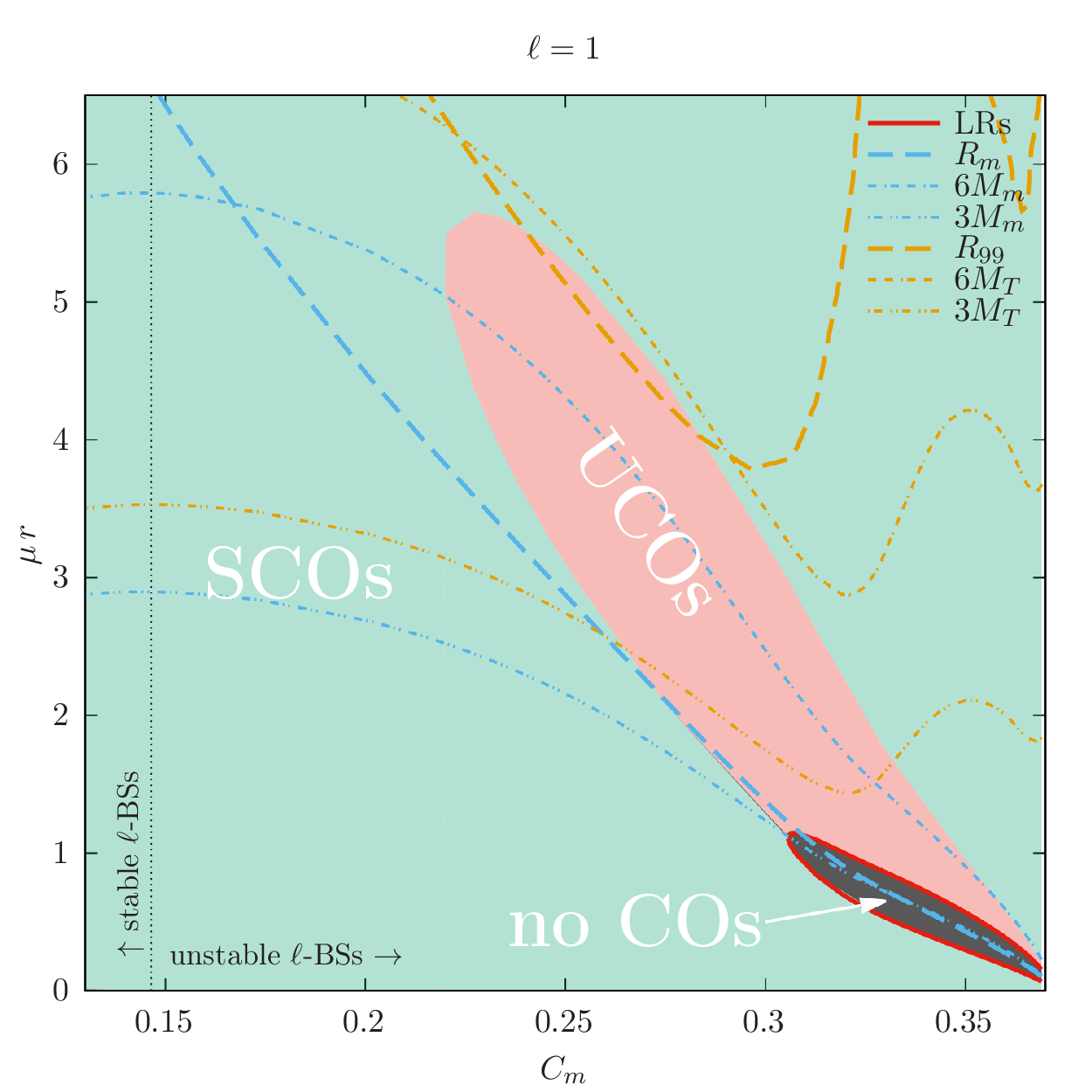}
\includegraphics[width=0.44\textwidth]{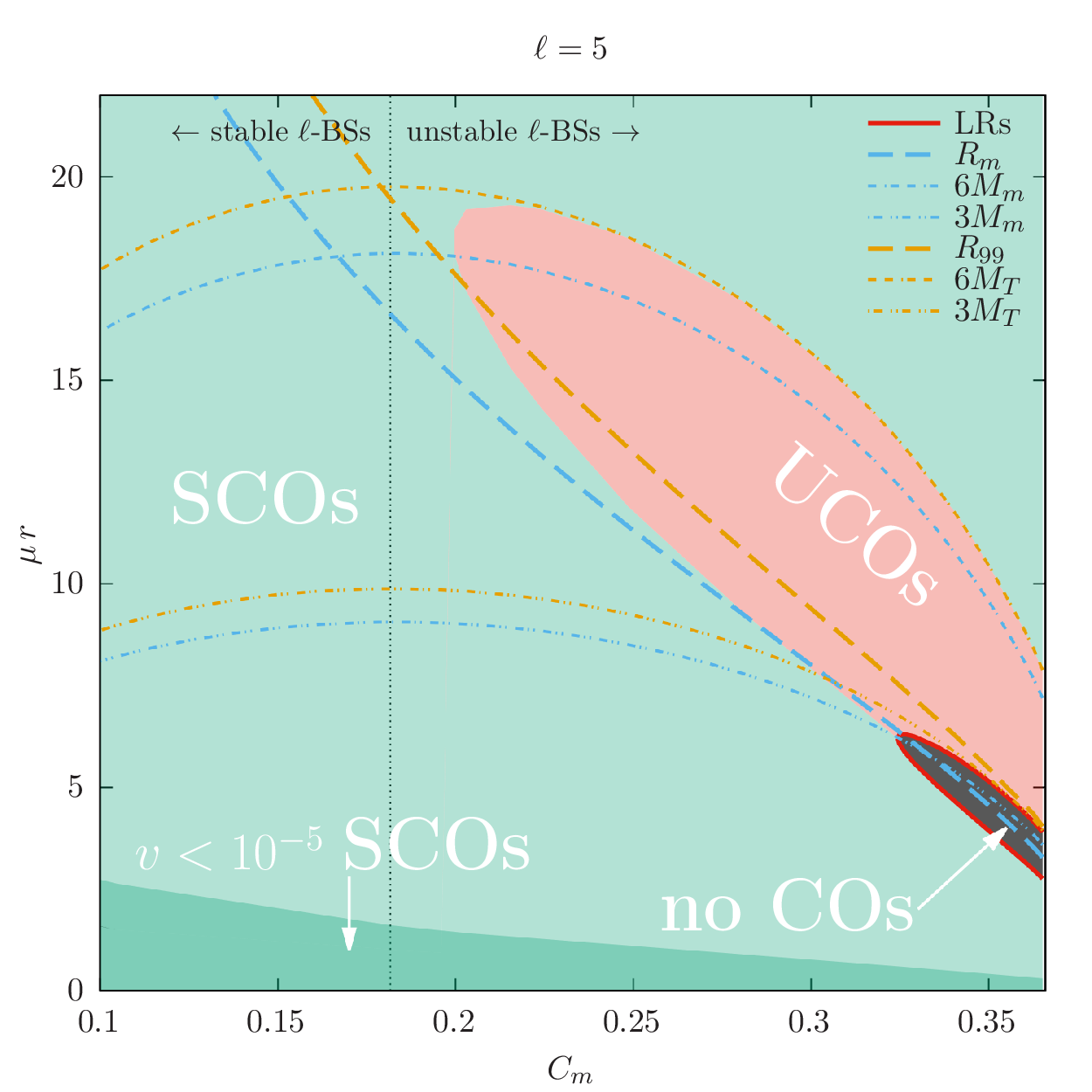}
\includegraphics[width=0.44\textwidth]{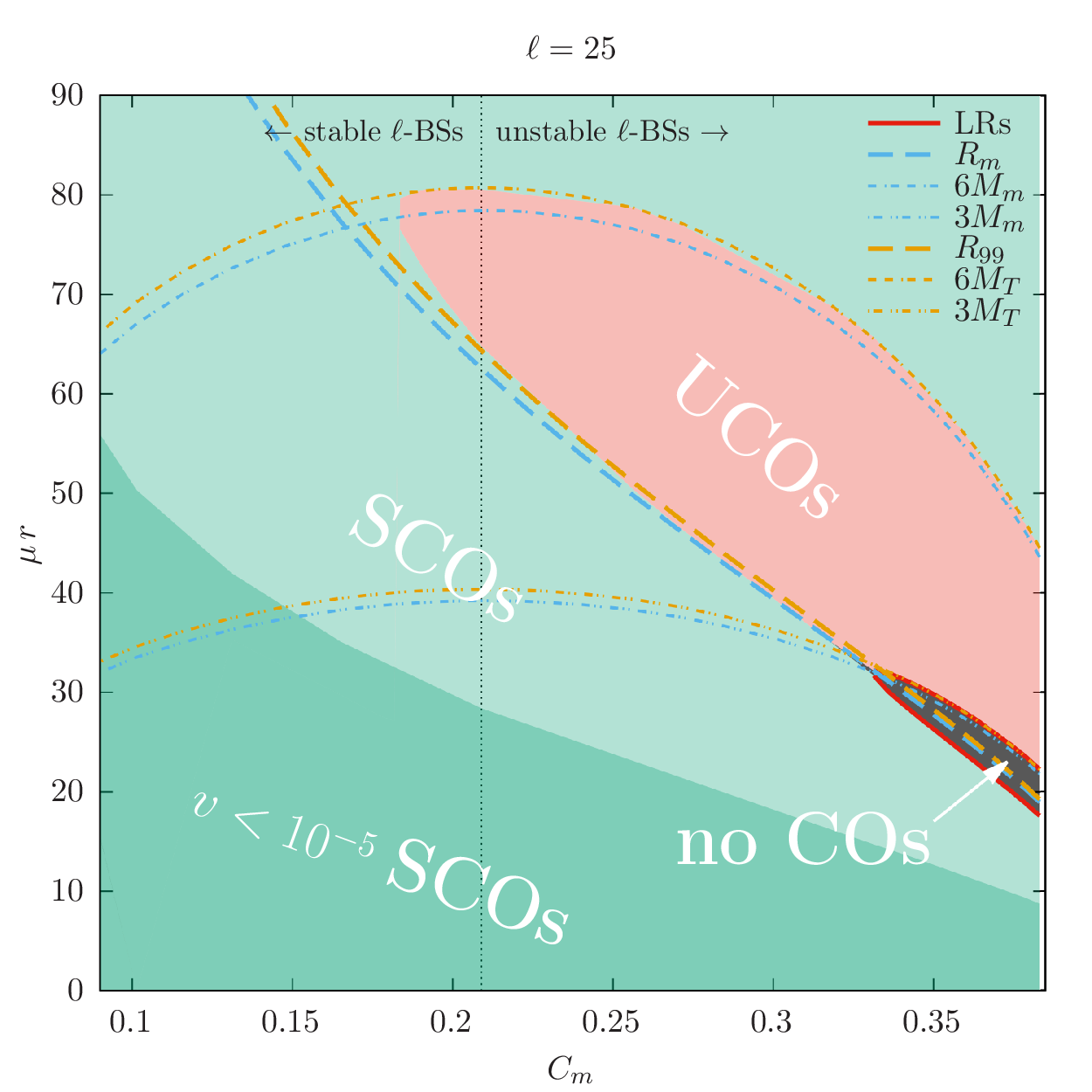}
\caption{For solutions of high compactness we indicate the regions of
  existence and stability of causal circular orbits (COs). In these plots, each vertical line of constant $C_m$ corresponds to a different solution.
   The green regions indicate the radii with stable (timelike) circular orbits
  (SCOs), while the red region indicates those with unstable orbits
  (UCOs). On the other hand, no COs exist in the dark gray region, which is limited by
  a pair of light rings (LRs), red line. Finally, the dark green region indicates the
  ``almost empty, almost flat'' central region of the $\ell>1$ ``shells'',
  where the circular orbits have speed $v<10^{-5}$. We also include
  as a guide $R_{99}$, $R_m$, and the corresponding locations of a
  Schwarzschild LR and ISCO.  
   }
 \label{LR_Cmax}
\end{figure}

\begin{figure}
\includegraphics[width=0.65\textwidth]{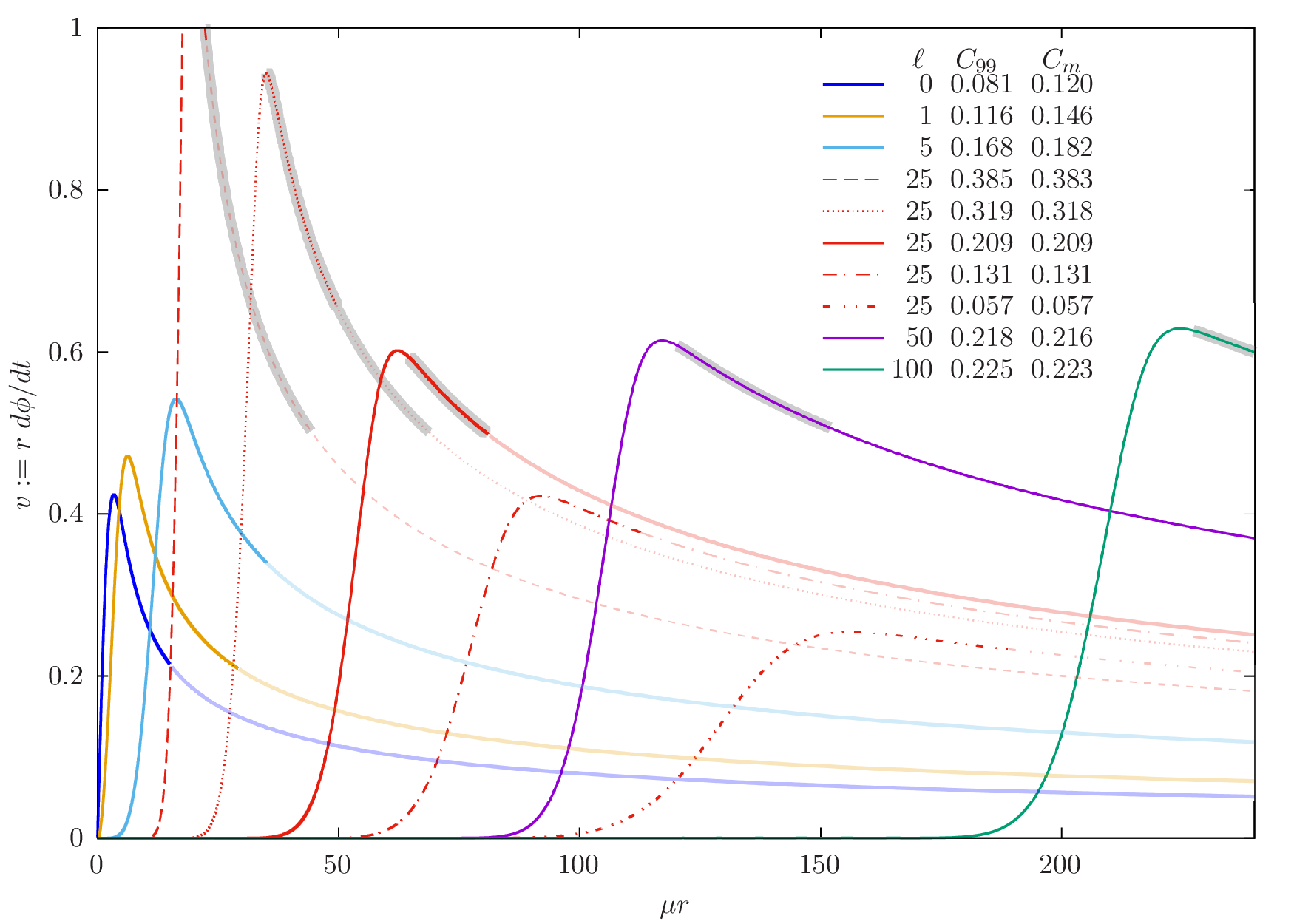}
\caption{Rotation curves as defined by equation~\eqref{eqRC}. The solid lines correspond to
  solutions with maximum $M_T$ for  $\ell$ ranging from $0$ to
  $100$. For $\ell=25$ we also show, in red, some cases with varying compactness
  around the maximum $M_T$ solution. 
  Beyond the numerical integration region, we extended the curves using the
  Schwarzschild spacetime with mass $M_T$. Those parts of the  curves are
  shown in a lighter color.
  Thick gray lines indicate regions where the  circular orbits are unstable.
  The solutions represented here are the
  same as those in figure~\ref{density}. 
   }
 \label{RC}
\end{figure}

We now go into more detail and analyze the different stability regions in
figure~\ref{LR_Cmax}, where we show plots of radius vs. compactness for
$\ell=0$, 1, 5 and 25. Each vertical line in these plots corresponds to a
solution, and we can see  the transitions through different
stability regions as $r$ varies along said line.
The green regions  are those where the timelike circular orbits are stable
(SCOs). The red region is where the circular orbits are unstable (UCOs), and
it is delimited by an ISCO at the top and by an OSCO at the bottom.
Similarly, the dark gray region is that for which no circular orbitss exist, and it is
delimited by a pair of light rings, indicated with a red line.
Since $\ell$-boson stars are shell like for $\ell>1$, with density much smaller than its maximum value
and falling quickly towards the center in the interior region
--where the spacetime is very close to Minkowski--
the circular orbits are almost non-existent there,
having speed $v\ll1$. To make this more apparent we shaded in a darker
green the regions in which $v<10^{-5}$, noting that the rotation curves of these
configurations have a maximum in the interval $0.4\lesssim v< 1$ (see
figure~\ref{RC}).

Figure~\ref{LR_Cmax} also indicates the stars' radii, $R_{99}$ and $R_m$, and
also, as a guide, the limit of spacetime stability (vertical dotted line) and
the locations of the Schwarzschild ISCO and light ring, given by $r=6M$ 
and $r=3M$, respectively, where for the value of $M$ we used both $M_T$ and
$M_m$.
We see that, as the compactness increases, the light rings first
appear at or very close to $R_m$, and soon they move to
each side of that location. 
For small $\ell$ the definition $R_m$ seems
more meaningful than $R_{99}$ when comparing to the location of light rings. On
the other hand, both definitions tend to coincide at large $\ell$.

Although all the solutions with light rings found in this
  work are unstable, we see that, for larger values of $\ell$, solutions with
  light rings exist closer and closer to the stable region. However, as mentioned
  earlier, it is unlikely that stable solutions with light rings
    exist, even for extremely large $\ell$.

Interestingly, the unstable circular orbit regions for large $\ell$ tend to be delimited
  almost exactly by the star's radius (from below) and the Schwarzschild ISCO
  (from above).
We can see again in the last panel of figure~\ref{LR_Cmax} ($\ell=25$) that regions of instability can exist
even for stable $\ell$-boson star spacetimes. As already mentioned, this
happens for solutions starting at $\ell=9$.
This could constitute an
observable feature that might help distinguish some $\ell$-boson stars from
other dark compact objects.

In figure~\ref{RC} we show the rotation curves for the solutions shown in
figure~\ref{density}, that is: solutions of maximum $M_T$ for $\ell=0$, 1, 5,
25, 50 and 100; and for $\ell=25$, also some solutions with varying compactness,
both in the stable and unstable spacetime branch. The curves have been extended beyond
the domain of numerical integration using the Schwarzschild expressions with
mass $M_T$. We can see that this gives and excellent match. 
The points where the curves reach $v=1$ correspond to light rings, and no circular
orbits exist in the region in
between those points (red line and dark gray region of
figure~\ref{LR_Cmax}). We also indicate the regions where the circular orbits are
unstable (thick gray line).

\section{Scaling properties for large \texorpdfstring{$\ell$}{l} \label{s:scaling}}

In this section we discuss the scaling properties of the fields in the asymptotic limit $\ell\to \infty$. This is achieved by rescaling the fields $(M, \alpha,\gamma,\psi)$ and by shifting and rescaling the radial coordinate $r$  in an appropriate way (which is largely motivated by the empirical numerical data and trial-and-error) such that, when taking the limit $\ell\to \infty$, one obtains a set of effective field equations which can be solved separately. As we show, combining the solution of these effective equations with the aforementioned rescaling, one obtains the correct asymptotic behavior for the fields and related quantities for large values of $\ell$. For clarity, we include a summary of these results in the final paragraph of this section.

\begin{figure}
  \includegraphics[width=0.42\textwidth]{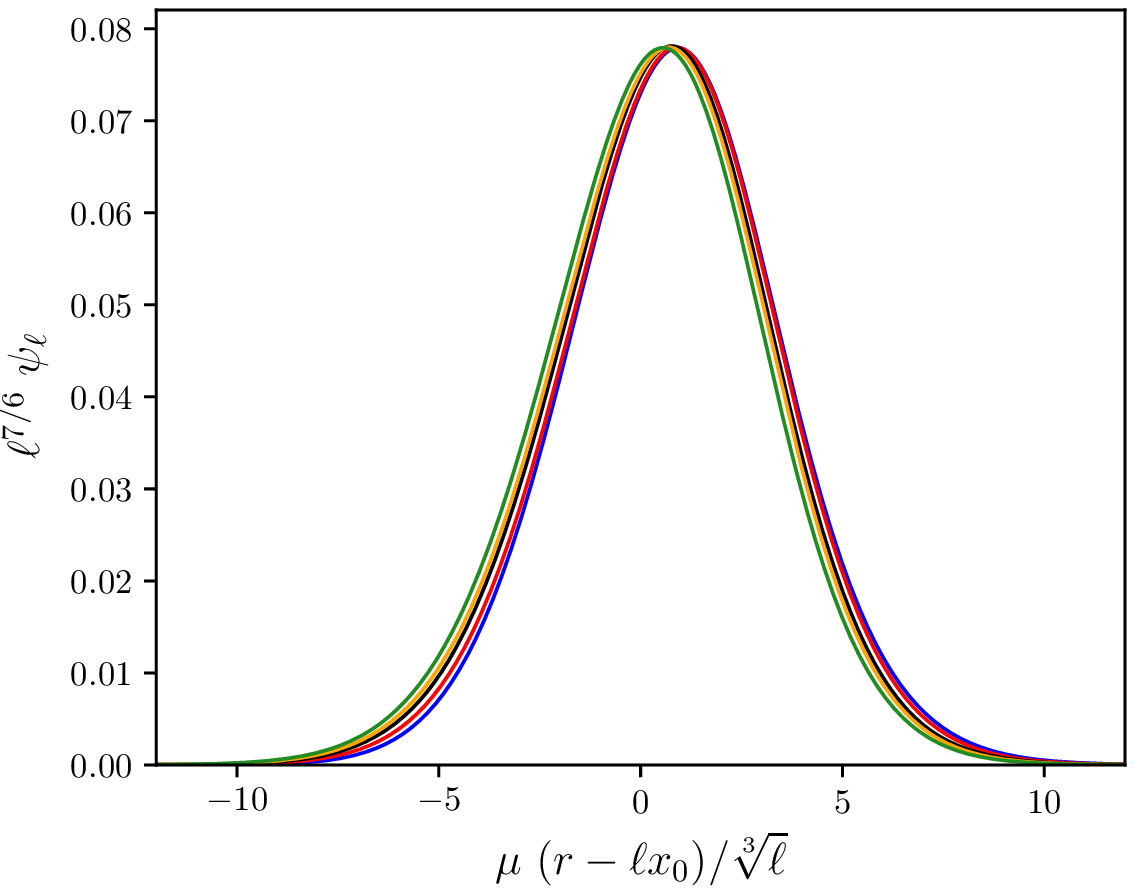}
  \includegraphics[width=0.42\textwidth]{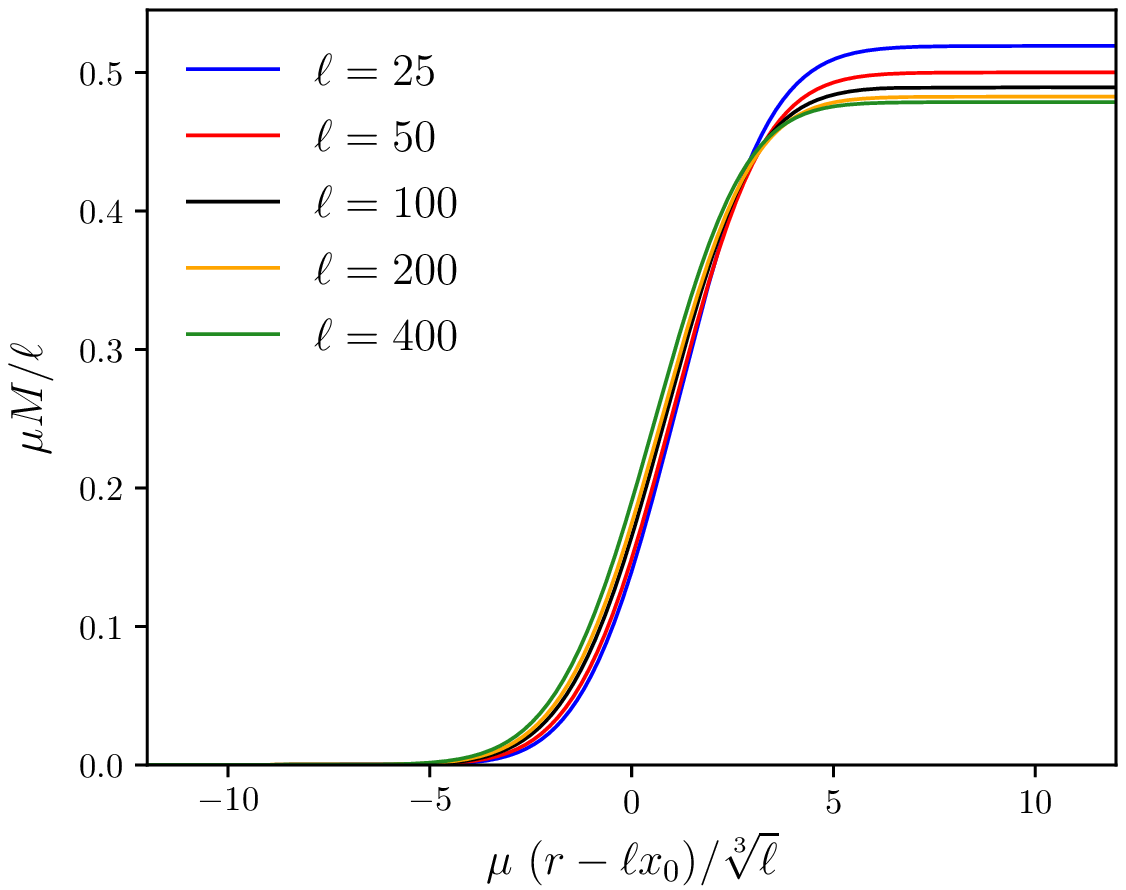}\\
  \includegraphics[width=0.42\textwidth]{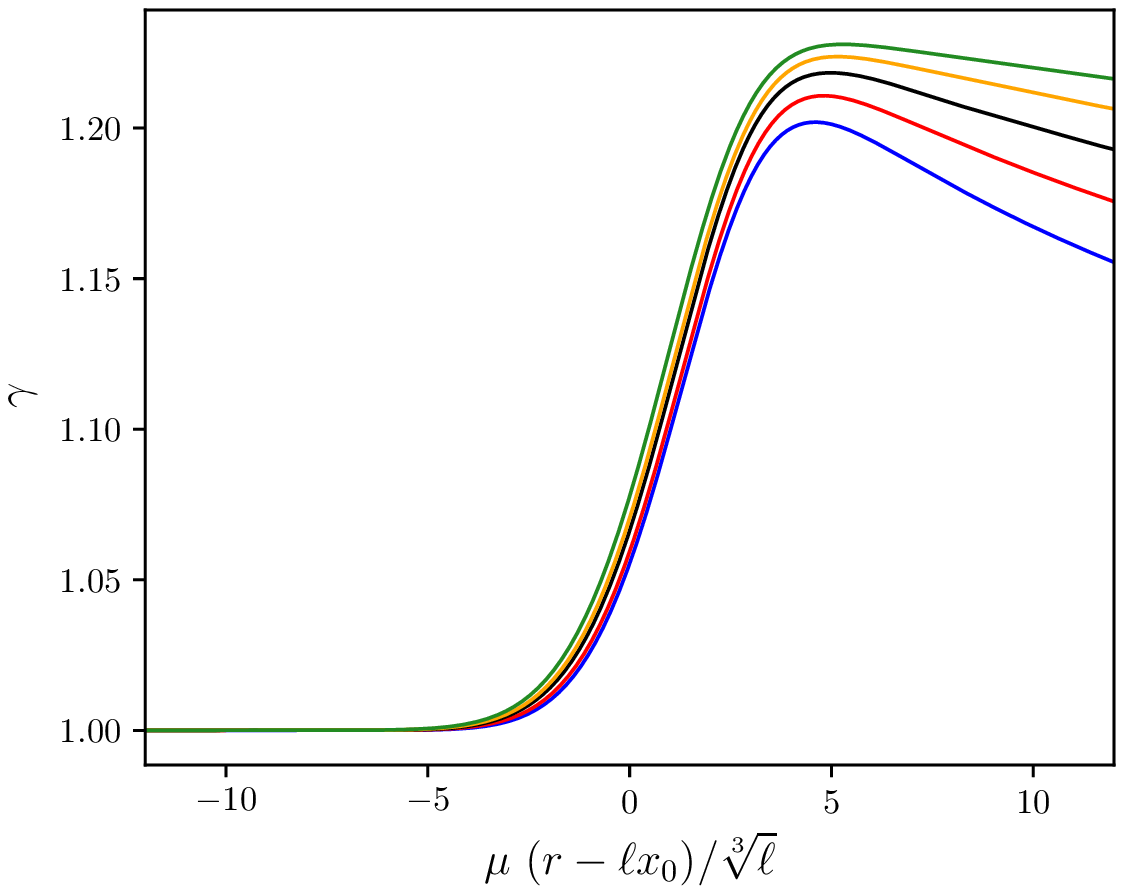}
  \includegraphics[width=0.42\textwidth]{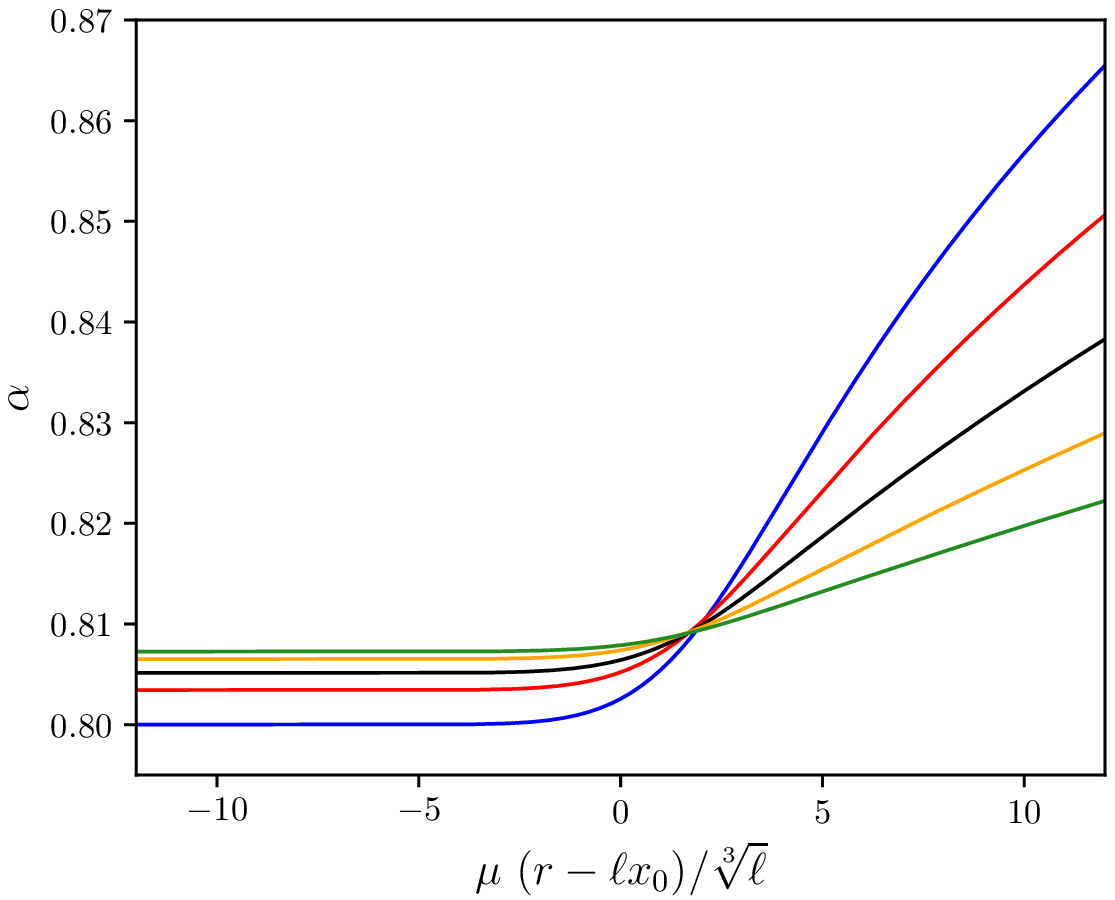}
\caption{Scaling of the solutions with $\omega=0.8612$, for the quantities $\psi_\ell$, $M(r)/\ell$, $\gamma$ and $\alpha$.
In these plots, the constant $x_0$ appearing on the horizontal axis has been estimated using as reference the $\ell=1600$ configuration with the same $\omega$,
using the formula $x_0=r_{1600}/1600$, with $r_{1600}$ defined as the location of the maximum of $\psi_{1600}(r)$. The procedure to determine $x_0$ without
resorting to any particular finite $\ell$ solution is explained in the text.
}
\label{ScalingPhi}
\end{figure}

To describe our scaling method, we consider a family of configurations with increasing value of $\ell$ and fixed $\omega$. As $\ell$ becomes large, the numerical data (see figure~\ref{ScalingPhi}) suggests that the fields' profiles depend only on the variable
\begin{equation}
y := \frac{r - \ell x_0}{\ell^a},
\label{Eq:yDef}
\end{equation}
with $x_0$ a positive constant that depends on $\omega$ but not $\ell$, and $a$ a parameter within the range $0 < a < 1$ that will be determined later. This means that the profiles have their center shifted outwards by $\ell x_0$ and stretched by the factor $\ell^a$ as $\ell\to \infty$. The fields' amplitudes are rescaled as follows:
\begin{equation}
M_*(y) := \frac{M(r)}{\ell},\qquad
\alpha_*(y) := \alpha(r)\qquad
\gamma_*(y) := \gamma(r)\qquad
\psi_*(y) := \ell^{1 + \frac{a}{2}}\psi_\ell(r),
\label{Eq:RescaledQuantities}
\end{equation}
the data suggesting that the quantities with a star have finite limits when $\ell\to \infty$. Note that equations~(\ref{Eq:yDef},\ref{Eq:RescaledQuantities}) and the definition of $\gamma$ in equation~(\ref{Eq:Metric}) imply that
\begin{equation}
\gamma_*^{-2}(y) = 1 - \frac{2M_*(y)}{x_0}\frac{1}{1 + \ell^{a-1}\frac{y}{x_0}},
\end{equation}
such that in the limit $\ell\to \infty$ (with fixed $y$) it follows that $\gamma_*^{-2}(y) = 1 - 2M_*(y)/x_0$. In terms of the rescaled quantities $M_*$, $\alpha_*$ and $\psi_*$ equations~(\ref{Eq:bosonstars}) can be written as
\begin{subequations}\label{Eq:bosonstarsRescaled}
\begin{eqnarray}
&& \frac{dM_*}{dy} = x_0^2\left( 1 + \ell^{a-1}\frac{y}{x_0} \right)^2\rho_*,
\label{Eq:bosonstarsRescaled.2} \\
&& \frac{1}{\gamma_*^2\alpha_*}\frac{d\alpha_*}{dy} 
 = x_0\left( 1 + \ell^{a-1}\frac{y}{x_0} \right) p_{r*} 
 + \frac{M_*}{x_0^2}\frac{\ell^{a-1}}{\left( 1 + \ell^{a-1}\frac{y}{x_0} \right)^2},
 \label{Eq:bosonstarsRescaled.3} \\
&& \frac{1}{\alpha_*\gamma_*} \frac{d}{dy}\left( \frac{\alpha_*}{\gamma_*} \frac{d\psi_*}{dy} \right)
 + \frac{2}{x_0}\frac{\ell^{a-1}}{1 + \ell^{a-1}\frac{y}{x_0}} \frac{1}{\gamma_*^2}\frac{d\psi_*}{dy}
 = -\ell^{2a}\left[  \frac{\omega^2}{\alpha_*^2} - \mu^2  - \frac{1}{x_0^2}\frac{1 + \frac{1}{\ell}}{\left( 1 + \ell^{a-1}\frac{y}{x_0} \right)^2} \right]\psi_*,
\label{Eq:bosonstarsRescaled.1}
\end{eqnarray}
\end{subequations}
where we have introduced the rescaled energy density and radial pressure
\begin{subequations}
\begin{eqnarray}
\rho_*(y) &:=& 4\pi\ell^{1+a}\rho(r) =  \left( 1 + \frac{1}{2\ell} \right)\left[
 \ell^{-2a}\frac{1}{\gamma_*^2}\left( \frac{d\psi_*}{dy} \right)^2 
 + \left( \frac{\omega^2}{\alpha_*^2} + \mu^2 + \frac{1}{x_0^2}\frac{1 + \frac{1}{\ell}}{\left( 1 + \ell^{a-1}\frac{y}{x_0} \right)^2} \right)\psi_*^2 \right],
\\
p_{r*}(y) &:=& 4\pi\ell^{1+a}p_r(r) = \left( 1 + \frac{1}{2\ell} \right)\left[
 \ell^{-2a}\frac{1}{\gamma_*^2}\left( \frac{d\psi_*}{dy} \right)^2 
 + \left( \frac{\omega^2}{\alpha_*^2} - \mu^2 - \frac{1}{x_0^2}\frac{1 + \frac{1}{\ell}}{\left( 1 + \ell^{a-1}\frac{y}{x_0} \right)^2} \right)\psi_*^2 \right].
\label{Eq:prRescaled}
\end{eqnarray}
\end{subequations}

Let us consider the limiting case $a=0$ first and take the limit $\ell\to \infty$ in these equations (with $y$ held fixed). In this case, one obtains the effective equations
\begin{subequations}\label{Eq:bosonstarsRescaleda=0}
\begin{eqnarray}
&& \frac{dM_\infty}{dy} = x_0^2\rho_\infty,\qquad
\rho_\infty = \frac{1}{\gamma_\infty^2}\left( \frac{d\psi_\infty}{dy} \right)^2 
 + \left( \frac{\omega^2}{\alpha_\infty^2} + \mu_0^2 \right)\psi_\infty^2,
\\
&& \frac{1}{\gamma_\infty^2\alpha_\infty}\frac{d\alpha_\infty}{dy} 
 = x_0 p_{r\infty},\qquad
p_{r\infty} = \frac{1}{\gamma_\infty^2}\left( \frac{d\psi_\infty}{dy} \right)^2 
 + \left( \frac{\omega^2}{\alpha_\infty^2} - \mu_0^2 \right)\psi_\infty^2,
\label{Eq:bosonstarsRescaled.3a=0}\\
&& \frac{1}{\alpha_\infty\gamma_\infty} \frac{d}{dy}\left( \frac{\alpha_\infty}{\gamma_\infty} \frac{d\psi_\infty}{dy} \right)
  = -\left( \frac{\omega^2}{\alpha_\infty^2} - \mu_0^2 \right)\psi_\infty,
\label{Eq:bosonstarsRescaled.1a=0}
\end{eqnarray}
\end{subequations}
where the index $\infty$ refers to the (pointwise) limit for $\ell\to\infty$, i.e. $M_\infty(y) = \lim_{\ell\to \infty} M_*(y)$ and similarly for $\alpha_\infty$, $\gamma_\infty$ and $\psi_\infty$. We have also introduced the shorthand notation $\mu_0 := \sqrt{\mu^2 + 1/x_0^2}$ in order to abbreviate the notation. Equations~(\ref{Eq:bosonstarsRescaleda=0}) look like a nice system of differential equations for $(M_\infty,\alpha_\infty,\psi_\infty)$ which could be integrated numerically and whose solution with the appropriate boundary conditions should approximate the solution of the full system when $\ell$ is large and $|y| \lesssim \ell$. However, it is not difficult to show that these equations imply that
\begin{equation}
\alpha_\infty\gamma_\infty p_{r\infty} = \textrm{const},
\end{equation}
and by virtue of the boundary conditions this constant must be zero. Therefore, $p_{r\infty} = 0$ which implies that $\alpha_\infty$ is constant and $\omega^2/\alpha_\infty^2 - \mu_0^2 < 0$. Then, multiplying both sides of equation~(\ref{Eq:bosonstarsRescaled.1a=0}) with $\psi_\infty$, integrating over $y$ and using integration by parts reveals that $\psi_\infty = 0$ is the only solution which decays to zero as $y\to \pm\infty$. This indicates that the choice $a = 0$ in the rescaling~(\ref{Eq:yDef}) is not the correct one.

Therefore, let us assume that $0 < a < 1$ is strictly positive and take again the pointwise limit $\ell\to \infty$ in equations~(\ref{Eq:bosonstarsRescaled}). This yields
\begin{subequations}\label{Eq:bosonstarsRescaleda>0}
\begin{eqnarray}
&& \frac{dM_\infty}{dy} = x_0^2\rho_\infty,\qquad
\rho_\infty = \left( \frac{\omega^2}{\alpha_\infty^2} + \mu_0^2 \right)\psi_\infty^2,
\label{Eq:bosonstarsRescaled.2a>0}\\
&& \frac{1}{\gamma_\infty^2\alpha_\infty}\frac{d\alpha_\infty}{dy} 
 = x_0 p_{r\infty},\qquad
p_{r\infty} = \left( \frac{\omega^2}{\alpha_\infty^2} - \mu_0^2 \right)\psi_\infty^2,
\label{Eq:bosonstarsRescaled.3a>0}
\end{eqnarray}
\end{subequations}
while the rescaled Klein-Gordon equation~(\ref{Eq:bosonstarsRescaled.1}) implies that $p_{r\infty}$ must vanish in order for the right-hand side to be finite. It follows that
\begin{equation}
\alpha_\infty = \frac{\omega}{\mu_0}
\label{Eq:alphainf}
\end{equation}
is constant and that
\begin{equation}
 \frac{dM_\infty}{dy} = 2x_0^2\mu_0^2\psi_\infty^2 = 2(1 + \mu^2 x_0^2)\psi_\infty^2.
\label{Eq:MinfEq}
\end{equation}

The problem is that (so far) we have no differential equation for $\psi_\infty$. However, a differential equation for $\psi_\infty$ can be obtained by expanding the rescaled fields:
\begin{equation}
\psi_*(y) = \psi_\infty(y) + \varepsilon\psi_1(y) + {\cal O}(\varepsilon^2),
\end{equation}
and similarly for $M_*$ and $\alpha_*$ in powers of $\varepsilon = \varepsilon(\ell)$ and looking at the next-order contributions from equations~(\ref{Eq:bosonstarsRescaled}). For the following, we choose $\varepsilon(\ell) = \ell^{a-1}$ since most of these corrections terms are of this order, and we expand the rescaled lapse in the form
\begin{equation}
\alpha_*(y) = \alpha_\infty\left[ 1 + \ell^{a-1}\delta(y) + {\cal O}(\ell^{-1}) \right],
\label{Eq:alphaExp}
\end{equation}
with the function $\delta(y)$ describing the first-order correction. Using equation~(\ref{Eq:alphainf}) the right-hand side of equation~(\ref{Eq:bosonstarsRescaled.1}) gives, to leading order in $1/\ell$,
\begin{equation}
2\ell^{3a-1}\left( \mu_0^2\delta - \frac{y}{x_0^3} \right)\psi_\infty,
\end{equation}
which yields a finite contribution if $a = 1/3$. Choosing $a = 1/3$ in the expansion~(\ref{Eq:alphaExp}), equations~(\ref{Eq:bosonstarsRescaled.3},\ref{Eq:bosonstarsRescaled.1}) yield the two differential equations
\begin{subequations}
\label{Eq:deltapsiinf}
\begin{eqnarray}
\frac{1}{\gamma_\infty^2}\frac{d\delta}{dy} 
 &=& x_0\left[ \frac{1}{\gamma_\infty^2}\left( \frac{d\psi_\infty}{dy} \right)^2 
  - 2\left( \mu_0^2\delta - \frac{y}{x_0^3} \right)\psi_\infty^2 \right] 
  + \frac{M_\infty}{x_0^2},
\label{Eq:deltainf}\\
\frac{1}{\gamma_\infty}\frac{d}{dy}\left( \frac{1}{\gamma_\infty}\frac{d\psi_\infty}{dy} \right) 
 &=&  2\left( \mu_0^2\delta - \frac{y}{x_0^3} \right)\psi_\infty,
\label{Eq:psiinf}
\end{eqnarray}
\end{subequations}
which can be integrated along with equation~(\ref{Eq:MinfEq}) and $\gamma_\infty^{-2} = 1 - 2M_\infty/x_0$ in order to find $(M_\infty,\delta,\psi_\infty)$. Note that the expression inside the square parenthesis on the right-hand side of equation~(\ref{Eq:deltainf}) is the $\ell^{-2/3}$-contribution to $p_{r*}$.

\begin{figure}
  \includegraphics[height=0.42\textwidth]{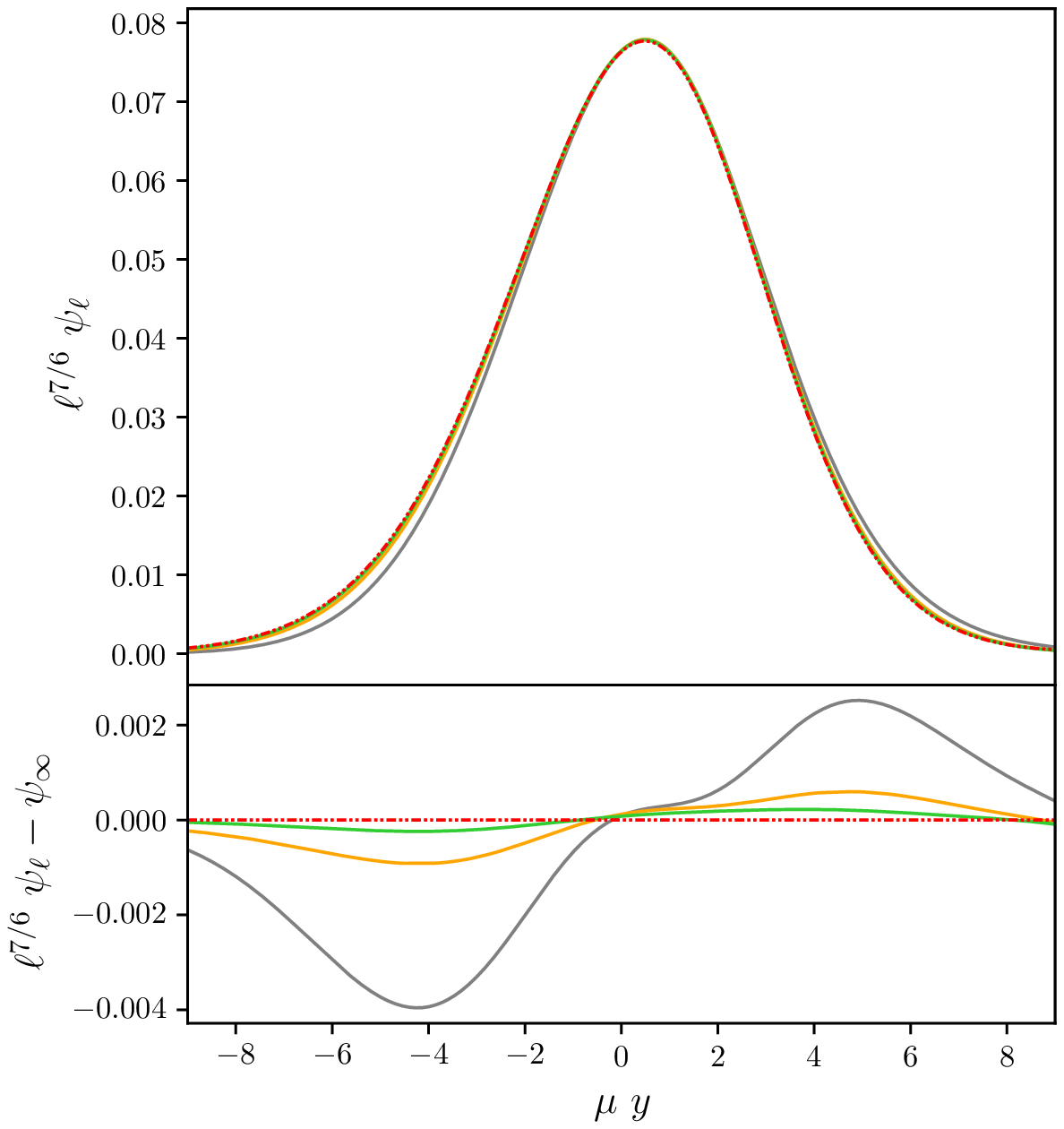}
  \includegraphics[height=0.42\textwidth]{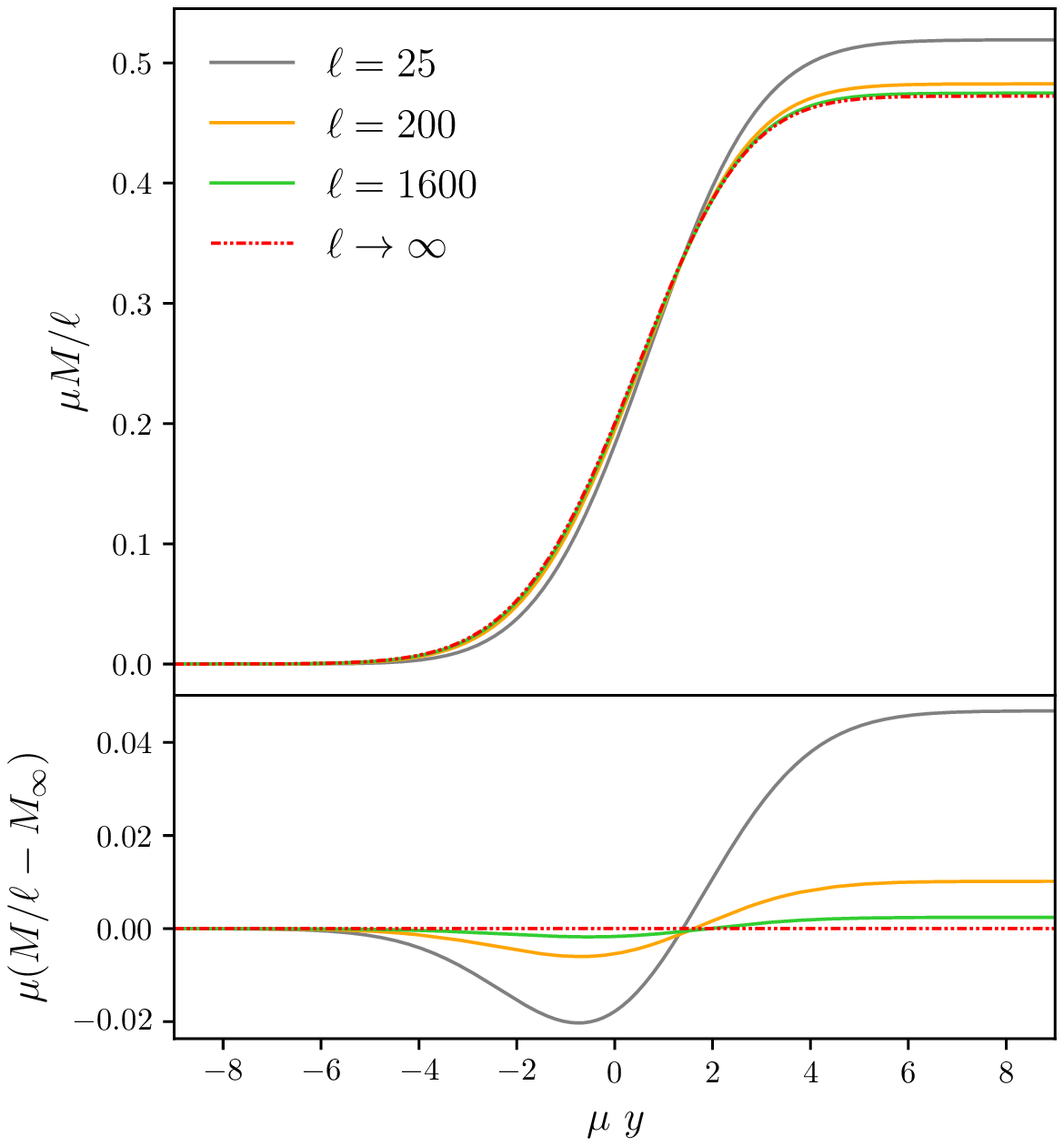}\\
\caption{Solutions with $\ell\gg 1$ compared to the same $\omega=0.8612$ solution obtained from the effective ($\ell\to\infty$ limit) equations~(\ref{Eq:MinfEq},\ref{Eq:deltapsiinf}). The asymptotic solution $\ell\to \infty$ yields the value $x_0 = 2.73$.
In the bottom panels we
show the difference between the finite $\ell$ configurations and the
$\ell\to\infty$ case, which converges to zero.
}
\label{ScalingFields}
\end{figure}

\begin{figure}
\includegraphics[width=0.32\textwidth]{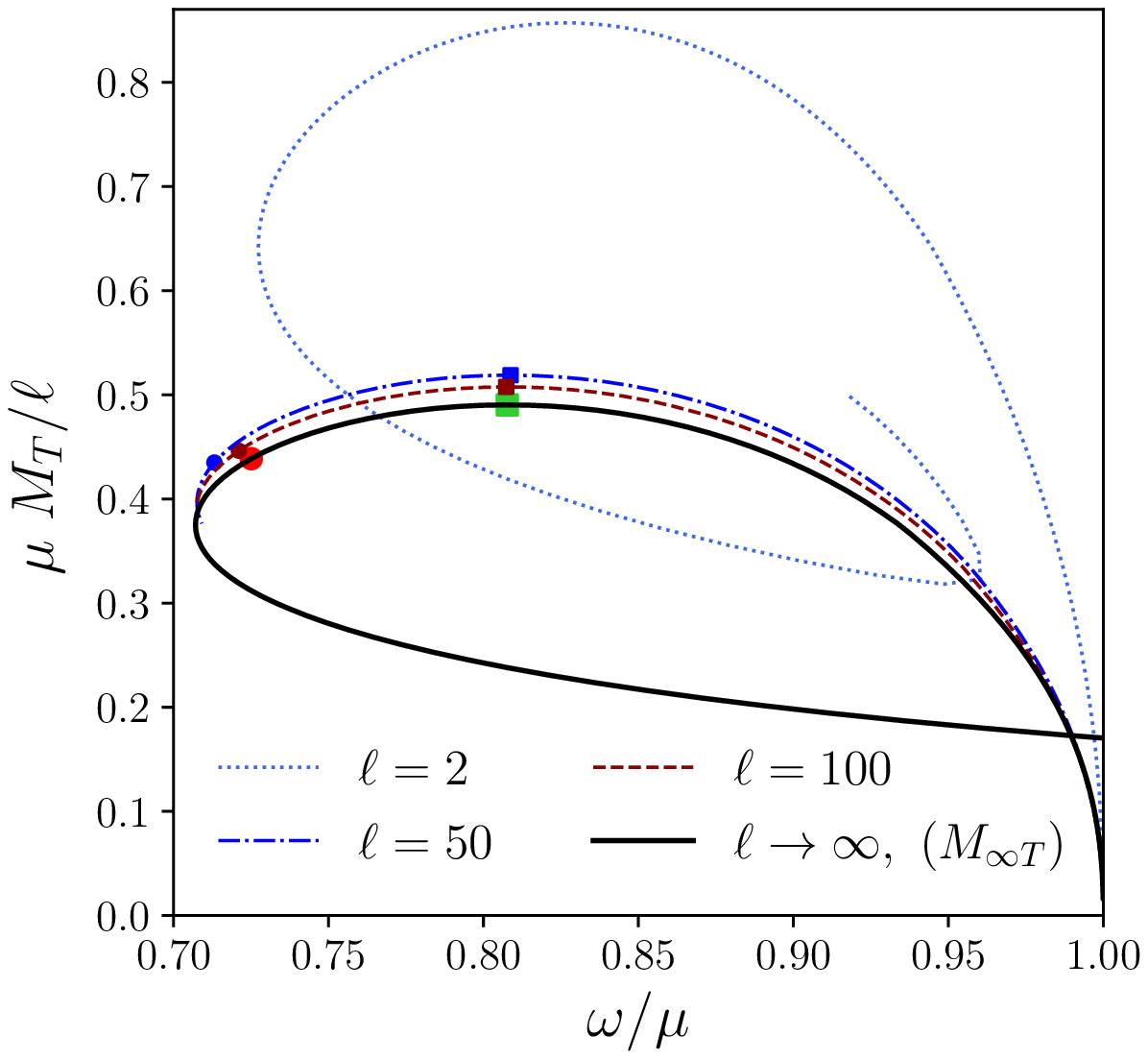} \includegraphics[width=0.323\textwidth]{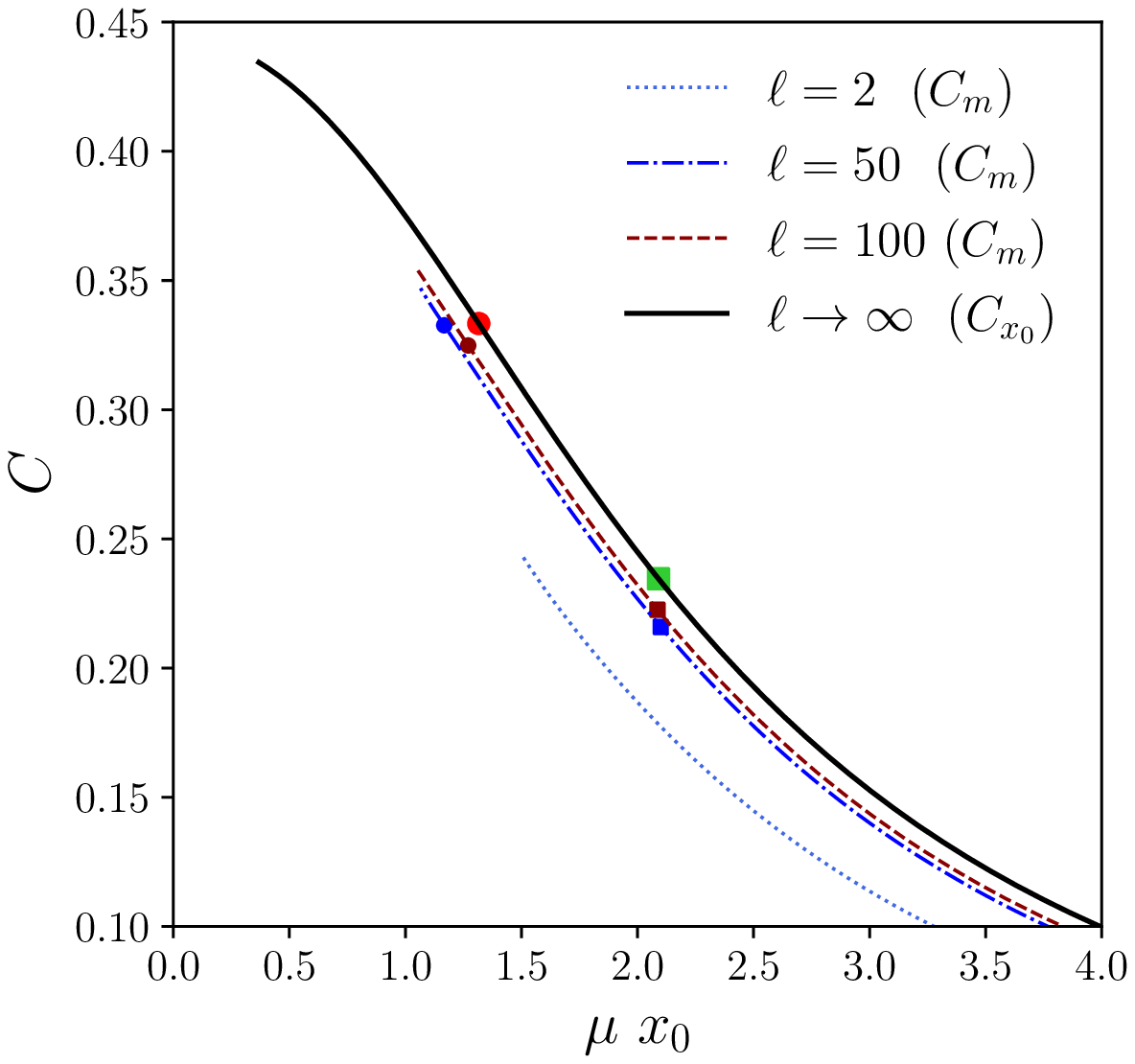} \includegraphics[width=0.3185\textwidth]{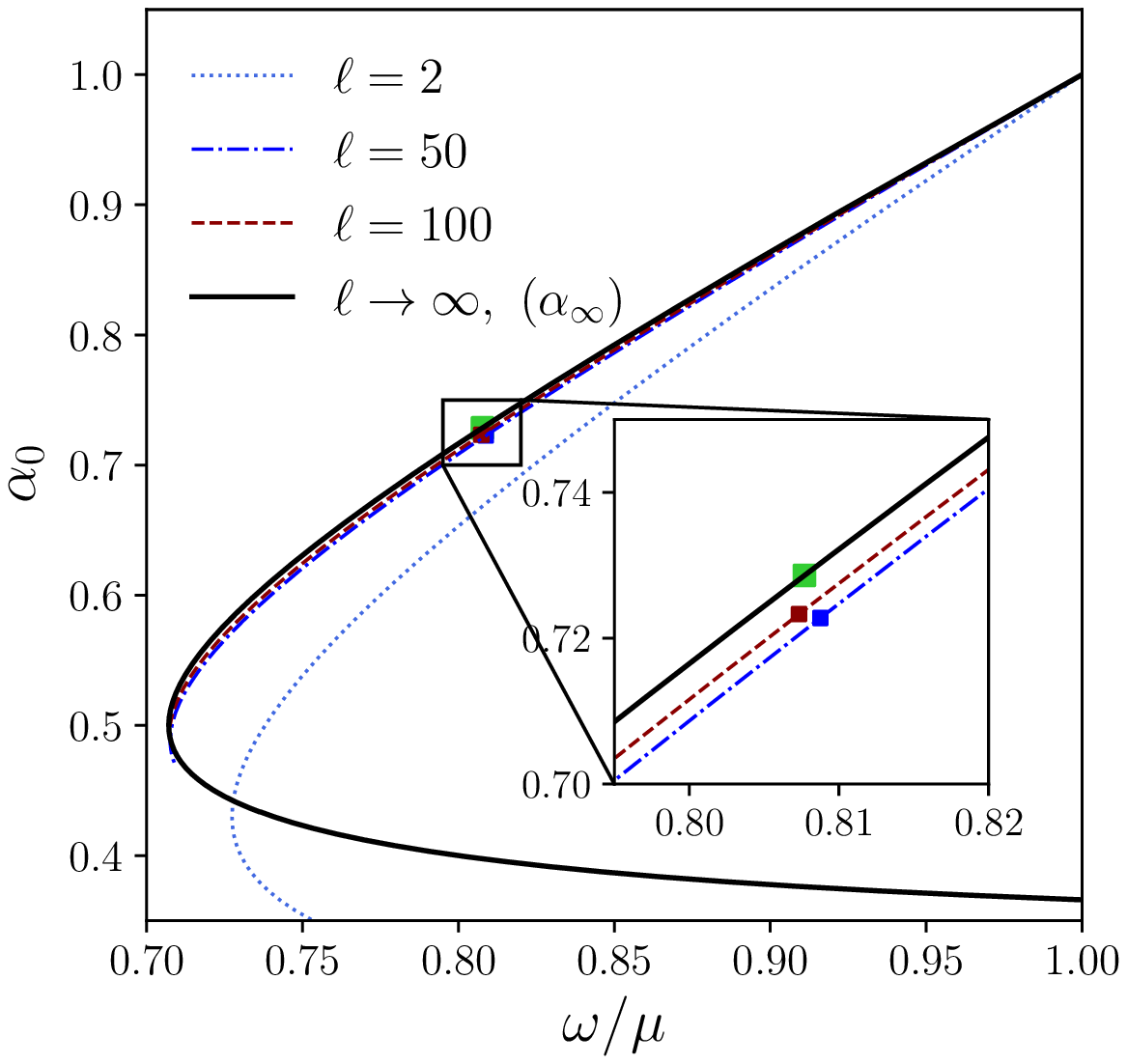}\\
\caption{Equilibrium $\ell\gg1$ configurations. Left panel: Total mass \textit{vs.} the frequency $\omega/\mu$. 
  Center panel: Compactness $C_{x_0}$ for the $\ell\to\infty$ limit. 
  Right panel: Minimum of the lapse $\alpha_0=\alpha(r=0)$ \textit{vs.} $\omega/\mu$. 
  The green square represents the maximum ($\ell\to\infty$) configuration and the red circle denotes the first appearance of light rings, at $C_{x_0}=1/3$. }
\label{Scaling2}
\end{figure}

The rescaled equations~(\ref{Eq:MinfEq},\ref{Eq:deltapsiinf}) are solved on a finite interval $[y_L,y_R]$ with $y_L<0<y_R$, 
fixing the left boundary conditions $M_*=0$, $\delta=0$, $\psi_*=\psi_{*L}$ at $y=y_L$ and
the right boundary condition $\psi_*\sim0$ at $y=y_R$. The integration is carried out by 
means of a shooting method from left to right, where $\psi_*$ is fixed at $y_L$
and the value of $x_0$ for which the field matches the boundary condition at $y_R$ is searched for.
In this procedure, it is necessary to provide the value of $d\psi_*/dy$ at $y_L$ given $\psi_{*L}$;
this can be done by studying the asymptotic behavior $y\to-\infty$ of the rescaled equations.
It is obtained that the scalar field takes the form $\psi_*\propto \mathrm{Ai}(z)\approx\exp(-\frac{2}{3}z^{3/2})/z^{1/4}$
with $z=-\sqrt[3]{2}y/x_0$ and $\mathrm{Ai}$ the Airy function of the first kind, obtaining 
$d\psi_*/dy\approx(\sqrt{-2y/x_0^3}-1/(4y))\psi_*$.

After $x_0$ is found, the total mass of the solution is obtained by evaluating $M_{\infty T}:=M_\infty(y=y_R)$. Outside the spherical shell, at $y=y_R$, we evaluate $\alpha_\infty=1/\gamma_\infty(y_R)$ and calculate $\omega$ from equation~(\ref{Eq:alphainf}). Once the pair $(\omega,x_0)$ is obtained from the effective equations, we can compare the fields with those corresponding to the
same $\omega$ finite $\ell$ solutions. Notice that there is no loss in generality in choosing $\delta(y_L)=0$ since 
the system (\ref{Eq:MinfEq},\ref{Eq:deltapsiinf}) is invariant under the transformation $(y,\delta) \mapsto (y-\zeta,\delta-\zeta/(\mu_0^2 x_0^3))$ with $M_\infty$ and $\psi_\infty$ unchanged.
In fact, as the lower right panel of figure~\ref{ScalingPhi} shows, the $\delta$ correction to $\alpha$ is not zero in the inner shell region. 
In turn, the previous transformation will translate horizontally the scalar field profile. So there are two ways to calculate $\zeta$, the first is to take a solution with large $\ell$ and find the value of the $\delta$ correction within the shell, the second is to use the scalar field profile and make the maxima of $\psi_*$ of the large $\ell$ solution and the effective $\ell\to\infty$ solution to overlap; these two forms are equivalent.

We illustrate  the fields' rescaling in figure~\ref{ScalingFields}, showing
convergence to the limiting $\ell\rightarrow\infty$ case, as expected. We 
have found that the value of $x_0$ that corresponds to a solution with the 
frequency of the previous (stable branch) figure~\ref{ScalingPhi} configurations, $\omega=0.8612$, is
$x_0=2.73$. The estimated value for the $\alpha$ correction in this case 
is $\zeta=2.17$. It is found that for this $\zeta$ value, the maxima of $\psi_*$
overlap, as expected. Figure~\ref{Scaling2} shows a plot for certain global quantities of the equilibrium solutions of the rescaled $\ell\to\infty$ limit. The left panel shows that a critical mass solution, $M_{\infty
  T}=0.49031$, is obtained at $\omega=0.8077$ (marked with a green square). 
This solution is obtained for the values $x_0=2.0902$ and $\zeta=1.58$.

Next, we evaluate the anisotropy and compactness of this particular configuration. In contrast to the rescaled radial pressure~(\ref{Eq:prRescaled}), the rescaled tangential pressure
\begin{equation}
p_{T*}(y):=4\pi\ell^{1+a}p_T=\left(1+\frac{1}{2\ell}\right)\left[-\ell^{-2a}\frac{1}{\gamma_*^2}\left(\frac{d\psi_*}{dy}\right)^2+\left(\frac{\omega^2}{\alpha_*^2}-\mu^2\right)\psi_*^2\right],
\end{equation}
does not vanish in the pointwise $\ell\to\infty$ limit:
\begin{equation}
p_{T\infty}=\left(\frac{\omega^2}{\alpha_\infty^2}-\mu^2\right)\psi_\infty^2 = \frac{\psi_\infty^2}{x_0^2},
\end{equation}
which is consistent with the observations made in section~\ref{s:Anisotropy}.

\begin{figure}
\includegraphics[width=0.45\textwidth]{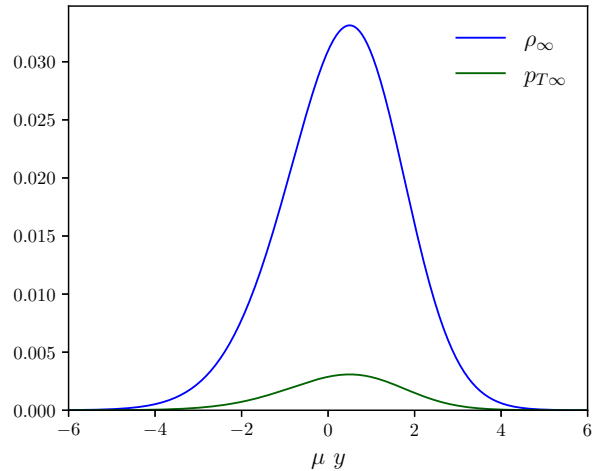}\\
\caption{Rescaled energy density and tangential pressure for the $\ell\to\infty$ maximum mass solution. Notice that $p_T$ is always positive, while $p_r$ is strictly zero in this limit.} 
\label{ScalingP}
\end{figure}

We show in figure~\ref{ScalingP} a plot for the tangential pressure as well as the rescaled energy density, equation~(\ref{Eq:bosonstarsRescaled.2a>0}),
for the solution of maximum mass. 
Now, to determine the compactness of these solutions, the easiest way is to note
that the quotient $M(r)/r$ in terms of the rescaled quantities in equations~(\ref{Eq:yDef},\ref{Eq:RescaledQuantities}), reduces to $M_\infty(y)/x_0$ in the $\ell\to\infty$ limit, allowing us to define,
\begin{equation}
C_{x_0}:=\frac{M_{\infty T}}{x_0}.
\end{equation}

In the central panel of figure~\ref{Scaling2} we show the $x_0$ value of the solution as a function of the compactness $C_{x_0}$ (and $C_m$ for the finite $\ell$ solutions).
For the maximum mass solution, the compactness obtained is $C_{x_0}=0.234554$. Like the finite $\ell$ solutions, the compactness increases as the value of the boson star radius decreases, approaching the limit value of $0.5$. However, the $\ell=\infty$ solutions with compactness exceeding $\simeq 0.433$ have frequencies $\omega$ larger than $\mu$, and thus they do not correspond to a limit of solutions with finite $\ell$ which must have $\omega/\mu <1$ due to the exponential decay of the scalar field at spatial infinity.

\begin{figure}
\includegraphics[width=0.45\textwidth]{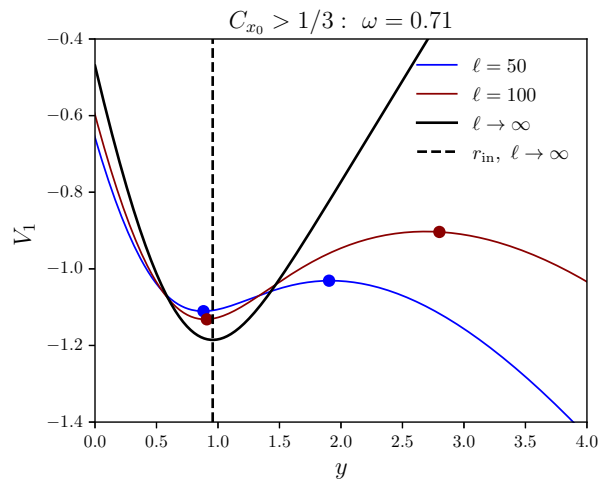}
\caption{Effective rescaled potential $V_1$ for circular null geodesics for $\ell=50,100$ solutions together with the $\ell\rightarrow\infty$ case. The circles indicate the
    position of the light rings.}
 \label{V1}
\end{figure}

Next, we wonder about the presence of light rings for the $\ell\gg1$ configurations. As stated above, the existence of these rings is given by the existence of local extrema of $V_\mathrm{eff}$. For large $\ell$ the effective potential for null geodesics is
\begin{equation}\label{Eq:proptoV}
V_\mathrm{eff}(r) = L^2\frac{\alpha^2}{r^2}
 = \frac{L^2}{\ell^2}\frac{\alpha_\infty^2}{x_0^2}\left[
 1 + \frac{2}{\ell^{2/3}}\left( \delta(y) - \frac{y}{x_0} \right) 
 + {\cal O}\left( \frac{1}{\ell} \right)
\right].
\end{equation}
For the following it is convenient to introduce the rescaled potential $V_1$, defined as
\begin{equation}
V_1(y) :=\ell^{2/3
}\left( \frac{x_0^2}{\alpha_\infty^2}\frac{\ell^2}{L^2} V_\mathrm{eff}(r)  - 1 \right)
 = 2\left( \delta(y) - \frac{y}{x_0} \right) + {\cal O}\left( \frac{1}{\ell^{1/3}} \right).
\end{equation}
Figure~\ref{V1} shows the function $V_1(y)$ for $\ell=50$ and $\ell=100$ along with the $\ell\rightarrow\infty$ limit. Starting from the knowledge of $\delta(y)$ we can obtain the approximate location of $r_\mathrm{in}$, the inner light ring for $\ell\gg1$ solutions. However, as shown in figure~\ref{V1} at this order one is unable to determine the position of the outer ring which moves away from $y=0$ as $\ell$ increases; in fact, it seems that the location of this second light ring diverges to $y\to \infty$ when $\ell\to \infty$.
Evaluating the condition $dV_1/dy=0$ along the $\ell\rightarrow\infty$ family of configurations we obtain that the solution closest to the critical mass point satisfying this condition for some value of $y$, is the one that has compactness $C_{x_0}=1/3$ (red dot in figure~\ref{Scaling2}). This correspond to the solution with $x_0=1.32$ and $\omega=0.7251$. 

To close the discussion of this section on the rescaling properties for large $\ell$, we list 
the transformations involving certain relevant quantities mentioned in the previous paragraphs. To do this, suppose the situation in which an $\ell$-boson star solution has been obtained for a certain value of $\omega$ and sufficiently large $\ell=\ell_1$; then starting from it we can obtain an approximate solution with arbitrarily large $\ell=\ell_2 > \ell_1$ for the same $\omega$ in the following way: first,
identify the position $r$ of the maximum of $\psi_\ell$ and estimate\footnote{The error induced by this estimation as well as the following ones presented in this paragraph is of the order $\ell_1^{-2/3}$.}
$x_0 =r/\ell_1$. Then, apply the transformation $r\mapsto \sqrt[3]{\ell_2/\ell_1}(r-\ell_1 x_0)+\ell_2 x_0$. As function of this redefined coordinate $r$, the amplitude of the scalar field becomes smaller according to $\psi_{\ell_1}(r)\mapsto\psi_{\ell_2}(r) = (\ell_1/\ell_2)^{7/6}\psi_{\ell_1}(r)$ while the mass function grows as $M(r)\mapsto (\ell_2/\ell_1) M(r)$. The energy density and the tangential pressure both decrease according to  $\rho(r)\mapsto (\ell_1/\ell_2)^{4/3}\rho(r)$ and $p_T(r)\mapsto(\ell_1/\ell_2)^{4/3}p_T(r)$. On the other hand, the mass, radius and compactness parameters rescale as follows: 
$(M_T,R_{99},C_{99})\mapsto (\ell_2/\ell_1 M_T,\ell_2/\ell_1 R_{99}, C_{99})$ and
$(M_m,R_m,C_m)\mapsto (\ell_2/\ell_1 M_m,\ell_2/\ell_1 R_m, C_m)$. Let this example above serve as an illustration of the rescaling properties; certainly a better way to obtain a solution 
with $\ell=\ell_2$ is to solve the effective equations and then obtain the quantities with finite $\ell$ by inverting the definitions of the rescaled variables. In this case the error would be of order $\ell_2^{-2/3}$ or even smaller for some of the quantities.


\section{Conclusions}
\label{Sec:Conclusions}

We have studied various properties of the recently introduced $\ell$-boson
stars~\cite{Alcubierre:2018ahf}, analyzing in each case the role played by the angular momentum parameter $\ell$,
and paying particular attention to the large $\ell$ regime.
These objects are
composed of $2\ell+1$ massive complex scalar fields and present
notable characteristics which single them out from the standard
$\ell=0$
boson stars, while still sharing
with them several common features
\cite{Alcubierre:2018ahf,Alcubierre:2019qnh,Alcubierre:2021mvs,Jaramillo:2020rsv,Sanchis-Gual:2021edp}.
Among these features
are the fact that both are formed with complex scalar fields
on a spherically symmetric spacetime;
they both admit diluted and compact solutions; and they possess
stable and unstable branches separated by the solution of maximum mass for
a given $\ell$.
On the other hand, we had previously~\cite{Alcubierre:2018ahf,Alcubierre:2019qnh,Alcubierre:2021mvs} observed some characteristics related to the $\ell$
parameter: an increase in the compactness and size of the maximum mass configurations and the fact that their morphology tends to form a hollow-like central region 
(even in the $\ell=1$ boson star case), with the position of the maximum of
density moving away from $r=0$.
The purpose of the present work was to take a step forward and perform a thorough examination of how these features change with $\ell$.
In particular, using different numerical methods, we were able to increase 
notably the
magnitude
of the parameter $\ell$ and finally, with the information of $\ell\gg1$
solutions
and a  careful analysis of the system of equations, we were
able to
study the limiting case when the parameter $\ell$ goes to infinity.

One of the interesting features that can be observed is the fact that for $\ell>1$ the density in the central region is much smaller than in the shell region.
We have shown in this work that, as $\ell$
grows, so does the object and also the almost empty central region, tending to form shells of scalar fields where the size of the almost empty central region is much larger than the size of
the region where the scalar field is mainly distributed. This tendency in the behavior goes all the way to infinity, making the objects look like larger and larger shells.
We have shown that, when $\ell\gg1$ the scalar field
profile is shifted outwards proportionally to $\ell$ while its width grows
as $\sqrt[3]{\ell}$.
Furthermore, the spatial components of the stress energy-momentum tensor  tend to be
highly anisotropic as $\ell$ increases. Indeed, as $\ell$
grows the radial pressure tends to zero, while the tangential one remains
finite. In this way, for
large values of $\ell$, the shells tend to have no radial pressure and
are supported solely by the tangential ones, analogous to the way in which a
Roman arch supports its own weight.
This increase in the anisotropy seems related to an increase in the compactness \cite{Dev:2000gt}
of the $\ell$-boson star. 
The mass of the solutions that divide the stable and the unstable branches, as
well as their size, grows with $\ell$, but  in such a way that the compactness
tends to a finite value. We have proven that in the $\ell\rightarrow\infty$ limit
the compactness tends to about $0.23$ for the maximum mass
configuration; that is, about half the Buchdahl limit. 
However, unstable configurations may be much more compact, 
reaching a compactness
of about $0.433$ in the large $\ell$ limit. In this regard, it is 
interesting to point out that (single and multiple) shell-type configurations 
have also been found when analyzing the spherically symmetric steady-state 
solutions of the Einstein-Vlasov system~\cite{Andreasson:2006ja}. In particular, 
it has been proven that such shells satisfy the Buchdahl inequality and that 
static shells of Vlasov matter can have $M(r)/r$ arbitrarily close to 
$4/9$~\cite{Andreasson:2006ab,Andreasson:2007ck}.

Regarding orbiting particles, the high compactness that $\ell$-boson
stars can achieve while remaining stable gives rise to new features, which differentiate
them from standard boson stars and also from black holes. Schwarzschild black holes have an ISCO located at $6M$, with no stable circular orbits below that value. Consequently, accretion disks around non-rotating black holes typically have an inner boundary and
 ``end'' at $r=6M$. {\em Stable} standard boson stars do not have ISCOs, meaning that they could in principle possess an accretion disk extending
all the way to the star's center. On the other hand, {\em stable} $\ell$-boson stars
exist with an ISCO-OSCO pair. In this case, accretion disks could show
a ``gap'' between the ISCO and OSCO, to then again extend all the way to the
center. These differences could constitute an important observable feature.

Besides causal circular orbits, we  studied null ones, also known as
  light rings. We found that, for each $\ell$, a pair of light rings appears
  at high enough compactness, the exterior one beeing unstable and the
  interior one stable. These light rings are always in the unstable spacetime
  regions, although they begin appearing closer and closer to the stable region as
  $\ell$ increases, which seems reasonable given that more compact stable
  solutions exist for larger $\ell$.
  Our findings are consistent with the results of~\cite{Cunha_2017}: 
  if a regular compact object has a light ring, it must have at least
  two\footnote{Except, of course, for the degenerate case in which the two
    light rings coincide.}, one of them
  being stable; and the presence of the stable light ring is expected to lead to nonlinear spacetime instabilities.

In our vast parameter exploration we have not
  included excited modes (higher frequency
solutions containing one or more nodes of $\psi_\ell$), which would be
unstable if results for standard boson stars also hold here~\cite{Balakrishna_1998}.
However, solutions that combine a stable ground state solution with
excited ones might again be stable~\cite{Bernal_2010}. We expect to address
these questions in future works.

Apart from the properties discussed in this article,
$\ell$-boson stars 
open up the possibility to consider
a larger landscape of solutions such as the ones described in 
\cite{Sanchis-Gual:2021edp}.
These results along with 
the existence of a stable branch for the $\ell$-boson stars
\cite{Alcubierre:2019qnh, Alcubierre:2021mvs, Jaramillo:2020rsv} make us conclude that
such localized bosonic systems may play an important role in modeling astrophysical objects, such as galactic halos or black hole mimickers with potential observable consequences. Further work along these lines is underway and will be presented in the near future.


\acknowledgments
We would like to thank H\r{a}kan Andr\'easson for discussions and pointing out to us the analogy between spherical steady-state collisionless gas configurations and $\ell$-boson stars.
This work was partially supported by DGAPA-UNAM through grants 
IN110218, IN105920, by CONACyT 
``Ciencia de Frontera" Projects
No.~304001 ``Estudio de campos escalares con aplicaciones en 
cosmolog\'ia y astrof\'isica" and
No.~376127 ``Sombras, lentes y ondas 
gravitatorias generadas por objetos compactos astrof\'isicos", and
by the European Union's Horizon 2020 research and innovation (RISE)
program H2020-MSCA-RISE-2017 Grant No. FunFiCO-777740.
ADT was partially supported by CONACyT grant CB-286897.
OS was partially supported by a CIC grant to Universidad Michoacana 
de San Nicol\'as de Hidalgo. VJ acknowledge financial support from CONACyT 
graduate grant program.

\appendix

\section{Definitions and rescaling in \texorpdfstring{$\mu$}{mu}}
\label{tables}
We include tables that provide summarized information in a single place, aiding in the reading of this article. A summary of the main definitions used in this work is shown in
Table~\ref{notation}. Rescaling rules in $\mu$, which allows one to obtain solutions for arbitrary $\mu$ from
the solution of any given $\mu_0$,  are shown in Table~\ref{rescaling}.
\begin{table}[H]\centering
  \caption{Summary  of the main definitions used in this article. \label{notation}}
  \begin{tabular}{l l l}
    \hline
    \hline
    Symbol          & Definition                                      &  Depends on  \\
    \hline
    \hline
    $M$             & Mass function, also $M(r)$                      & $\ell$, $u_0$, $r$\\
    \hline
    $M_T$           & Total mass (or mass function at outer boundary) & $\ell$, $u_0$\\
    $R_{99}$        & Areal radius containing 99\% of the total mass  & $\ell$, 
$u_0$\\
    $C_{99}$        & $M_T/R_{99}$                                    & $\ell$, $u_0$\\
    \hline
    $C_m$           & Maximum of $M(r)/r$ over $r > 0$                            & $\ell$, $u_0$\\
    $R_m$           & Location of maximum $M(r)/r$                    & $\ell$, $u_0$\\
    $M_m$           & Mass function evaluated at $R_m$                & $\ell$, $u_0$\\
    \hline
    $r_{\rm in}$    & Location of the inner light ring                & $\ell$, $u_0$\\
    $r_{\rm out}$   & Location of the outer light ring                & $\ell$, $u_0$\\
    $r_{\rm osco}$  & Location of the OSCO                            & $\ell$, $u_0$\\
    $r_{\rm isco}$  & Location of the ISCO                            & $\ell$, $u_0$\\
    \hline
    $M_{\rm max}$   & Maximum of $M_T$ (for a given $\ell$)           & $\ell$\\
    \hline
    \hline
  \end{tabular}
\end{table}

\begin{table}[H]\centering
  \caption{Solutions for arbitrary values of $\mu$ can be obtained from those
    of a given value by performing a rescaling as shown in this table. 
    \label{rescaling}}
  \begin{tabular}{r c l}
    \hline\hline
    $\mu$ & $\mapsto$            & $\lambda\, \mu$  \\
    \hline\hline
    ($\alpha$, $\gamma$, $\psi_\ell$) & $\mapsto$   & ($\alpha$, $\gamma$, $\psi_\ell$) \\
    $u_0$ & $\mapsto$              & $\lambda^{\ell}\, u_0$ \\
    $\omega$ & $\mapsto$         & $\lambda\, \omega$ \\
    ($r$, $M$) & $\mapsto$              & $\lambda^{-1}$ ($r$, $M$) \\
    ($\rho$, $p_r$, $p_T$) & $\mapsto$           & $\lambda^2$ ($\rho$, $p_r$, $p_T$) \\
    \hline \hline
  \end{tabular}
\end{table}

\section{Numerical methods}
\label{Sec:Methods}

We obtain solutions of the eigenvalue problem in equations~(\ref{Eq:bosonstars}) numerically using two different
methods, implemented in independent codes. For $\ell \lesssim 25$ we use a
shooting method similar to the one described in our previous work~\cite{Alcubierre:2018ahf}, but with some
improvements. For larger $\ell$ it becomes more and more difficult for this
code to converge to a given mode. In those cases we switch instead to a
spectral method.
These methods, which are described in the following subsections, give the same
results in the parameter region where both are able to obtain solutions.

\subsection{Shooting Method}

To obtain solutions for $\ell \lesssim 25$ we use a
``shooting to a fitting point method'' based on~\cite{Press86},
implemented in a code which is described in~\cite{Megevand:2007uy}.
It consist of doing a direct numerical integration of the ordinary differential equations
starting both from the left and right boundaries, at which one imposes either
appropriate physical conditions or guesses when those are undetermined, with the
goal of matching both the fields and their first derivatives at some
intermediate point. This defines a function of the mentioned guesses, plus an
additional guess, the eigenvalue $\omega^2$, whose roots correspond to  the
fitting condition being satisfied. In order to find such roots, a
Newton-Raphson method is used.
The fitting point method is particularly useful when one has a system with
a pair of solutions, one rapidly growing and the other rapidly decreasing at
each boundary, and one wants to obtain the (physical) solution that decays to
zero at both boundaries, as in the large $\ell$ cases.
For the numerical integration, instead of the algorithm described
in~\cite{Press86}, we use a more sophisticated step adaptive method provided by the
LSODE routines~\cite{radha1993}.
For the particular applications of this work, it was also helpful in a few
cases to
modify the left boundary conditions in order to be able to set them at
locations quite a bit to the right of $r=0$. This is due to the shell-like
shape of the stars for large enough $\ell$. The details are given below. 
Finally, even though  the solutions for a given value of $\mu$ can be
trivially obtained from a rescaling of the $\mu=1$  case (see appendix~\ref{tables}), which is the
value we fixed in most situations, sometimes it helped the numerical code to easily
find solutions to vary
$\mu$ depending on the particular region of the parameter space. This is
because some fields may become many orders of magnitude different when one
restricts oneself to the $\mu=1$ case. Nevertheless,
we present all our results in a $\mu$ independent form.

\subsubsection{Approximate solutions  for low density}
\label{smallrho}
As mentioned throughout the article, for large values of $\ell$ the scalar field distribution is shell-like, with
very low density (as compared to its maximum value) in an interior region with $r<r_1$ and in an exterior region with $r_2<r$ for certain values $r_1<r_2$. In the
interior region the solutions can be approximated by those of a scalar field on
a flat spacetime, while in the exterior region they can be approximated by
solutions of a scalar field on a Schwarzschild spacetime with mass $M_T$.

\begin{figure}
\includegraphics[width=0.42\textwidth]{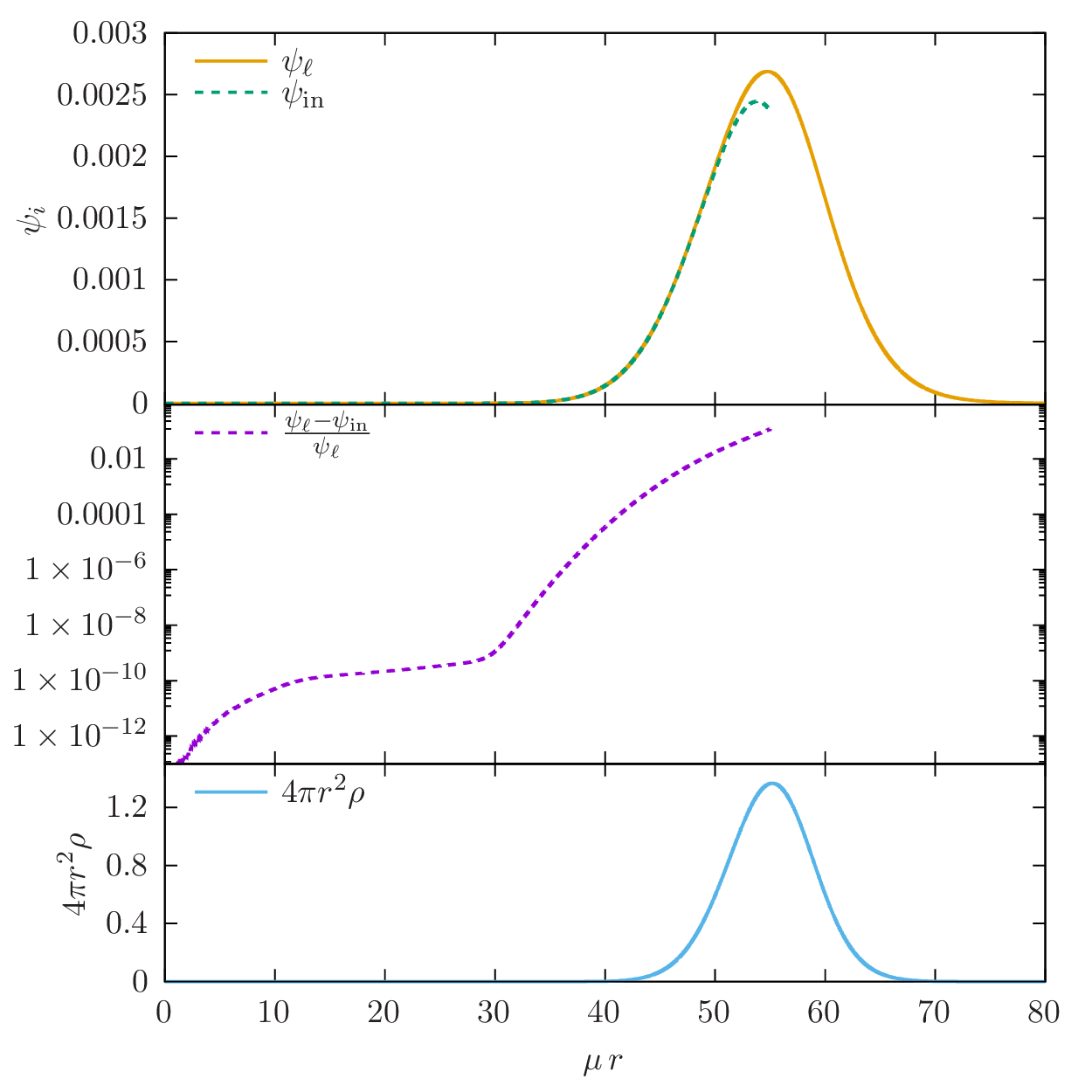}
\caption{Solution $\psi_\ell$ and approximation $\psi_\textnormal{in}$ for the case with
  $\ell=25$ and maximum mass $M_T=13.45$ (top panel). The middle panel shows
  the relative error. For reference we also include the density profile in the
bottom panel. }
\label{bessell25}
\end{figure}

In the interior region ($r<r_1$) we can assume $\alpha=\gamma=1$. Then, from equation~(\ref{Eq:bosonstars.1}), we get
\begin{equation}
\frac{1}{r^2}\left(r^2\psi_{\rm in}' \right)' 
 = \left(\mu^2 - \omega^2 + \frac{\ell(\ell+1)}{r^2} \right)\psi_{\rm in},  
\end{equation} 
with solutions
\begin{equation}
  \psi_{\rm in}(r) = C_1 \frac{ J_{\ell+\frac{1}{2}} \left( \sqrt{\omega^2-\mu^2}\, r    \right) }{ \sqrt{r} } +
            C_2 \frac{ Y_{\ell+\frac{1}{2}} \left( \sqrt{\omega^2-\mu^2}\, r    \right) }{ \sqrt{r} },
\end{equation}
where $J_\nu(x)$ and $Y_\nu(x)$ are the Bessel functions of the first and second kind,
respectively. Keeping only the solution with the proper behavior at $r=0$ and
writing the arbitrary amplitude in terms of $u_0$ we obtain
\begin{equation}
  \psi_{\rm in}(r) = u_0 \, \frac{ 2^{\left(\ell+\frac{1}{2}\right)} \, \Gamma\left(\ell+\frac{3}{2}\right) }
  {  {\left( \sqrt{\omega^2-\mu^2} \right)}^{\ell+\frac{1}{2} }   } \,
  \frac{ J_{\ell+\frac{1}{2}} \left( \sqrt{\omega^2-\mu^2}\, r    \right) }{ \sqrt{r} } .
\label{solin}
\end{equation}
In order to transform to the gauge used in the remainder of this work, in which
$\alpha=1$ at $r=\infty$, rather than at $r=0$, one just needs to replace
$\omega$ with $\gamma_\infty\,\alpha_\infty\,\omega$ in equation~(\ref{solin}).
We show an example of this approximation in figure~\ref{bessell25}. We see a
very good agreement between the scalar field and its approximation even well
beyond $\mu\,r=\mu\,r_1\approx 40$.

Finally, we note that in the exterior region one can assume the metric is given by a Schwarzschild
solution with mass $M_T$. Then, the scalar field can be expressed in
terms of the confluent Heun functions. However, we did not use the external
region approximations in this paper, hence we will not present any details here.

\subsection{Spectral Method}

An independent code was built based on a multidomain spectral method. 
Specifically, a collocation method has been used with Chebyshev polynomials
as the basis functions. Details of the code described in the following paragraphs
were essentially implemented based on~\cite{Grandclement:2007sb} which is a review 
on spectral methods in numerical relativity.
The Einstein-Klein-Gordon equations were solved
in isotropic coordinates, where the differential operators in the resulting equations
in the system are similar to each other and therefore easier
to implement in this particular method.

The physical domain, parametrized by the radial coordinate is decomposed into 5
carefully placed domains depending on what range of solutions we want
to obtain in a single run, given $\ell$. For the outer domain a compactification is carried out
so that the external boundary conditions can be imposed at spatial infinity.
On the other hand, in the domain that contains the origin, an even base
of Chebyshev polynomials is used for the lapse and the conformal factor $\Psi$, while
an even (odd) base is used for the field if $\ell$ is even (odd), which 
guarantee the solution is regular at the origin.
The non-linear system of equations that results for the coefficients of the
expansion is solved iteratively using a Newton scheme, the extra variable
$\omega$ is compensated with an extra equation $\alpha(r=0)=\alpha_0>0$, which
ensures that the code does not converge to the trivial solution.

An initial guess is required in the Newton scheme for the coefficients of
the expansion in all the functions (as well as the frequency), this is equivalent
to provide an initial guess for the functions. Given certain value of $\ell$, 
the
first solution is obtained from reasonable choices for the three 
parameters, $\sigma$, $r_0$ and $\phi_0$, which control
the properties of the following simple initial guess
\begin{align}
\psi_\ell&=\left(\frac{\bar{r}}{r_0}\right)^\ell \phi_0 \exp\left(-\frac{\bar{r}^2-r_0^2}{\sigma^2}\right),\\
\alpha&=-(1-\alpha_0)\exp(-\bar{r}^2)+1,\\
\Psi&=1.
\end{align}
Here $\bar{r}$ refers to the isotropic radial coordinate. The solutions are easier to find 
in the Newtonian regime where the frequency is close to one.
For example, in the $\ell=50,\ 100$ cases presented here we
start with an initial guess of $\omega=0.95$ and once the first solution is obtained
we slightly decrease the value of $\alpha_0$ and take as the new initial guess the
previous solution.

We have checked that in the spectral code, as in the convergence test performed in \cite{Grandclement:2014msa} 
for the $\ell=0$ case, the error indicators, as for example the frequency and the difference of the ADM and Komar 
masses converge exponentially to a fixed value and to zero respectively,
as we increase the number of Chebyshev basis polynomials, as expected for a spectral method.

\section{Summary of numerical data}
\label{table_data}
Table~\ref{data} shows information regarding most of the solutions analyzed in this paper. 
\begin{table}[H]\centering
  \caption{Properties of some of the solutions obtained in this work.
    For each $\ell$ (and $\mu$), the ground state solution is uniquely
    determined by fixing one more parameter, like $u_0$ or $\alpha_0$. We
    mainly use $u_0$ for small $\ell$ and $\alpha_0$ for large $\ell$. Hence,
    in most cases, we only report one of these parameters.
    In the column titled ``stable'', ``m.s.'' stands for marginally stable,
    corresponding to solutions of maximum total mass, which define the transition
    point between stable and unstable solutions. In the $\ell=\infty$ case
    (marked with ${}^\dag$), m.s. simply indicates the maximum mass
    solution. Also in the $\ell=\infty$ case, the total mass values (indicated
    with ${}^\star$) actually correspond to the mass rescaled with $\ell$ as
    in section~\ref{s:scaling}, since otherwise those values are not finite.}\label{data}
\scriptsize
\begin{tabular}{cccccccccccccccc}\hline\hline
  $\ell$   & $u_0/\mu^\ell$ & $\alpha_0$ & $\omega/\mu$ & $\mu M_T$     & $\mu M_m$ & $\mu R_{99}$ & $\mu R_m$ & $C_{99}$ & $C_m$  & $\mu r_{\rm in}$ & $\mu r_{\rm out}$ & $\mu r_{\rm osco}$ & $\mu r_{\rm isco}$ & stable        & Figure                                                                                                                     \\\hline\hline
  0        & 1.30e-2        &            & 0.9911       & 0.229         & 0.1707    & 41.908       & 22.332    & 0.0055   & 0.0076 & --               & --                & --                 & --                 & yes           & {\color{white}1,2,4,5,}\ref{prpt}{\color{white},8,9,10,11,12,14}                                                           \\
  0        & 2.71e-1        &            & 0.8530       & 0.633         & 0.4587    & 7.855        & 3.815     & 0.0806   & 0.1202 & --               & --                & --                 & --                 & m.s.          & {\color{white}1,}\ref{density},\ref{CmaxC99},\ref{pressures},\ref{prpt},\ref{RC}{\color{white},9,10,11,12,14}              \\
  0        & 2.20e+0        &            & 0.8428       & 0.374         & 0.0016    & 5.043        & 0.006     & 0.0742   & 0.2581 & 0.0031           & 0.0082            & 0.0031             & 0.013              & no            & {\color{white}1,2,4,5,}\ref{prpt}{\color{white},8,9,10,11,12,14}                                                           \\\hline
  1        & 5.00e-4        &            & 0.9864       & 0.489         & 0.4041    & 47.763       & 32.090    & 0.0102   & 0.0126 & --               & --                & --                 & --                 & yes           & \ref{Vrho},{\color{white}2,4,5,}\ref{prpt}{\color{white},8,9,10,11,12,14}                                                  \\
  1        & 4.00e-3        &            & 0.9487       & 0.875         & 0.7233    & 23.115       & 15.460    & 0.0379   & 0.0468 & --               & --                & --                 & --                 & yes           & \ref{Vrho}{\color{white},2,4,5,6,8,9,10,11,12,14}                                                                          \\
  1        & 3.35e-2        &            & 0.8353       & 1.176         & 0.9650    & 10.157       & 6.590     & 0.1158   & 0.1464 & --               & --                & --                 & --                 & m.s.          & \ref{Vrho},\ref{density},\ref{CmaxC99},\ref{pressures},\ref{prpt},\ref{RC}{\color{white},9,10,11,12,14}                    \\
  1        & 5.01e-1        &            & 0.8384       & 0.543         & 0.3739    & 3.860        & 1.224     & 0.1407   & 0.3056 & 1.1057           & 1.1191            & 1.1057             & 2.99               & no            & \ref{Vrho}{\color{white},2,4,5,6,8,9,10,11,12,14}                                                                          \\
  1        & 1.60e+0        &            & 0.8742       & 0.702         & 0.1504    & 7.333        & 0.430     & 0.0958   & 0.3500 & 0.2947           & 0.5691            & 0.2947             & 0.952              & no            & \ref{Vrho}{\color{white},2,4,5,6,8,9,10,11,12,14}                                                                          \\
  1        & 2.50e+0        &            & 0.8603       & 0.670         & 0.1008    & 6.183        & 0.280     & 0.1084   & 0.3596 & 0.1875           & 0.3936            & 0.1875             & 0.603              & no            & \ref{Vrho}{\color{white},2,4,5,6,8,9,10,11,12,14}                                                                          \\
  1        & 7.00e+0        &            & 0.8883       & 0.613         & 0.0376    & 6.579        & 0.102     & 0.0932   & 0.3690 & 0.0666           & 0.1542            & 0.0666             & 0.213              & no            & {\color{white}1,2,4,5,}\ref{prpt}{\color{white},8,9,10,11,12,14}                                                           \\\hline
  5        & 1.00e-10       &            & 0.9757       & 1.686         & 1.5377    & 72.612       & 61.560    & 0.0232   & 0.0250 & --               & --                & --                 & --                 & yes           & {\color{white}1,2,4,5,}\ref{prpt}{\color{white},8,9,10,11,12,14}                                                           \\
  5        & 5.00e-7        &            & 0.8165       & 3.293         & 3.0197    & 19.470       & 16.628    & 0.1691   & 0.1816 & --               & --                & --                 & --                 & m.s.          & {\color{white}1,}\ref{density},\ref{CmaxC99},\ref{pressures},\ref{prpt},\ref{RC}{\color{white},9,10,11,12,14}              \\
  5        & 3.40e-3        &            & 0.9016       & 1.314         & 1.1984    & 4.037        & 3.282     & 0.3255   & 0.3651 & 2.76             & 3.90              & 2.76               & 7.87               & no            & {\color{white}1,2,4,5,}\ref{prpt}{\color{white},8,9,10,11,12,14}                                                           \\\hline
  25       & 4.67e-53       &            & 0.9499       & 9.265         & 8.9616    & 162.304      & 156.420   & 0.0571   & 0.0573 & --               & --                & --                 & --                 & yes           & {\color{white}1,}\ref{density},{\color{white}4,5,6,}\ref{RC}{\color{white},9,10,11,12,14}                                  \\
  25       & 4.67e-47       &            & 0.8826       & 12.490        & 12.1047   & 95.484       & 92.363    & 0.1308   & 0.1311 & --               & --                & --                 & --                 & yes           & {\color{white}1,}\ref{density},{\color{white}4,5,}\ref{prpt},\ref{RC}{\color{white},9,10,11,12,14}                         \\
  25       & 1.03e-42       &            & 0.8091       & 13.451        & 13.0691   & 64.363       & 62.556    & 0.2090   & 0.2089 & --               & --                & 64.8               & 80.55              & m.s.          & {\color{white}1,}\ref{density},\ref{CmaxC99},\ref{pressures},\ref{prpt},\ref{RC},{\color{white}9,10,11,12,}\ref{bessell25} \\
  25       & 1.87e-36       &            & 0.7167       & 11.488        & 11.2098   & 35.994       & 35.251    & 0.3192   & 0.3180 & --               & --                & 35.055             & 68.85              & no            & {\color{white}1,}\ref{density},{\color{white}4,5,6,}\ref{RC}{\color{white},9,10,11,12,14}                                  \\
  25       & 9.35e-30       &            & 0.7608       & 7.425         & 7.2693    & 19.263       & 18.969    & 0.3855   & 0.3832 & 17.601           & 22.275            & 17.601             & 44.55              & no            & {\color{white}1,}\ref{density},{\color{white}4,5,}\ref{prpt},\ref{RC}{\color{white},9,10,11,12,14}                         \\
  25       & 1.08e-45       & 0.8        & 0.8612       & 12.980        & 12.5892   & 84.434       & 81.799    & 0.1537   & 0.1539 & --               & --                & --                 & --                 & yes           & {\color{white}1,2,4,5,6,8,}\ref{ScalingPhi},\ref{ScalingFields}{\color{white},11,12,14}                                    \\\hline
  50       & 1.00e-94       & 0.72277    & 0.8088       & 25.942        & 25.5042   & 119.187      & 118.120   & 0.2177   & 0.2159 & --               & --                & 120.1              & 152.0              & m.s.          & {\color{white}1,}\ref{density},\ref{CmaxC99},{\color{white}5,6,}\ref{RC}{\color{white},9,10,11,12,14}                      \\
  50       & 1.00e-99       & 0.80343    & 0.8612       & 25.004        & 24.5308   & 156.170      & 154.380   & 0.1601   & 0.1589 & --               & --                & --                 & --                 & yes           & {\color{white}1,2,4,5,6,}\ref{ScalingPhi},{\color{white}1,}\ref{Scaling2}{\color{white},11,12,14}                          \\\hline
  100      & 1.00e-216      & 0.72331    & 0.8073       & 50.756        & 50.2279   & 225.660      & 225.660   & 0.2249   & 0.2226 & --               & --                & 227.1              & 298.0              & m.s.          & {\color{white}1,}\ref{density},\ref{CmaxC99},{\color{white}5,6,}\ref{RC}{\color{white},9,10,11,12,14}                      \\
  100      & 1.00e-220      & 0.80513    & 0.8612       & 48.911        & 48.8731   & 296.630      & 296.630   & 0.1648   & 0.1631 & --               & --                & --                 & --                 & yes           & {\color{white}1,2,4,5,6,8,}\ref{ScalingPhi},{\color{white}10,}\ref{Scaling2}{\color{white},12,14}                          \\\hline
  200      &                & 0.8065     & 0.8612       & 96.510        & 95.8531   & 575.580      & 577.070   & 1.6767   & 0.1661 & --               & --                & --                 & --                 & yes           & {\color{white}1,2,4,5,6,8,}\ref{ScalingPhi},\ref{ScalingFields}{\color{white},11,12,14}                                    \\\hline
  400      &                & 0.80725    & 0.8612       & 191.452       & 190.6695  & 1128.450     & 1132.820  & 0.1697   & 0.1683 & --               & --                & --                 & --                 & yes           & {\color{white}1,2,4,5,6,8,}\ref{ScalingPhi}{\color{white},10,11,12,14}                                                     \\\hline
  1600     &                & 0.808      & 0.8612       & 759.695       & 758.6214  & 4422.320     & 4437.550  & 0.1718   & 0.1710 & --               & --                & --                 & --                 & yes           & {\color{white}1,2,4,5,6,8,9,}\ref{ScalingFields}{\color{white},11,12,14}                                                   \\\hline
  $\infty$ &                & 0.8086     & 0.8612       & $0.472^\star$ &           &              &           & 0.1730   &        &                  &                   &                    &                    &               & {\color{white}1,2,4,5,6,8,9,}\ref{ScalingFields},\ref{Scaling2}{\color{white},12,14}                                       \\
  $\infty$ &                & 0.7286     & 0.8077       & $0.490^\star$ &           &              &           & 0.2346   &        &                  &                   &                    &                    & m.s.${}^\dag$ & {\color{white}1,2,4,5,6,8,9,10,}\ref{Scaling2},\ref{ScalingP}{\color{white},14}                                            \\
  $\infty$ &                & 0.5773     & 0.7251       & $0.439^\star$ &           &              &           & 0.3333   &        &                  &                   &                    &                    &               & {\color{white}1,2,4,5,6,8,9,10,}\ref{Scaling2}{\color{white},12,14}                                                        \\\hline\hline
\end{tabular}
\end{table}

\afterpage{\clearpage}

\bibliographystyle{apsrev}
\bibliography{ref}


\end{document}